\documentclass[%
reprint,
amsmath,amssymb,
aps,xcolor, array,graphics, graphicx
]{revtex4-1}
\usepackage{xcolor}
\usepackage{graphicx}
\usepackage{dcolumn}
\usepackage{bm}
\usepackage{placeins}
\usepackage{float}
\usepackage{array}
\newcolumntype{L}{>{\centering\arraybackslash}m{2.5cm}}

\begin{document}
	
	
	\title{Chromodynamic multi-relaxation time lattice Boltzmann scheme for fluids with density difference }
	
	\author{J. Spendlove $^1$}
	\author{X. Xu $^{1,2}$}
	\author{O. J. Halliday $^{3}$}
	\author{T. Schenkel $^{1,2}$}
	\author{I. Halliday $^1$}
	
	\affiliation{$^1$ Materials \& Engineering  Research Institute, Sheffield Hallam University, Howard
		Street, S1 1WB (UK)}
	\affiliation{$^2$ Department of Engineering and Mathematics, Sheffield Hallam University, Howard
		Street, S1 1WB (UK)}
	\affiliation{$^3$ National Centre for Atmospheric Science, Department of Meteorology, University of Reading, Reading RG6 6AH (UK)}
	
	\date{\today}
	
	\begin{abstract}
		We develop, after Dellar ( P. J. Dellar, Phys. Rev. E. 65, 036309 (2002), J. Comput. Phys. 190, pp351 (2003)),
		a multiple-relaxation time (MRT), chromodynamic, multi-component
		lattice Boltzmann equation (MCLBE) scheme for simulation of isothermal, immiscible fluid flow with a density contrast. It is based on Lishchuk's method 
		(J. U. Brackbill, D. B. Kothe and C. Zemach, J. Comp. Phys. 100, 335-354 (1992), S. V. Lishchuk, C. M. Care and I. Halliday,  Phys. Rev. E. 67(3), 036701(2), (2003)) and 
		the segregation of d'Ortona et al. (U. D'Ortona, D. Salin, M. Cieplak, R. B. Rybka and J. R. Banavar Phys. Rev. E. 51, 3718, (1995)).
		We focus on fundamental model verifiability but do relate some of our data to that from previous approaches, due to Ba et al. 
		(Y. Ba, H. Liu, Q. Li, Q. Kang and J. Sun, Phys. Rev. E 94, 023310 (2016)) and earlier
		Liu et al. (H. Liu, A. J. Valocchi and Q. Kang, Phys. Rev. E 85, 046309 (2012)), who pioneered large density difference 
		chromodynamic MCLBE and showed the practical benefits of a MRT collision model.
		Specifically, we test the extent to which chromodynamic MCLBE MRT schemes 
		comply with the kinematic condition of mutual impenetrability and the continuous traction condition by developing analytical benchmarking flows.
		We conclude that our data, taken with those of Ba et al., verify the utility of MRT chromodynamic MCLBE.
	\end{abstract}

	\pacs{Valid PACS appear here}
	\maketitle
	
	%
	%
	%
	%

	\section{Introduction }
	Since 1991, when Gunstensen and Rothman \cite{Gunstensen} invented the technique, several
	multi-component lattice Boltzmann equation  (MCLBE) variants have developed to address different flow regimes 
	\cite{Succi, Kruger, Huang}. The idea remains a milestone of statistical physics,
	however all current MCLBE variants depart substantially from \cite{Gunstensen}, which developed directly 
	from Rothman's earlier immiscible lattice gas cellular automata, \cite{RothmanKeller, Roth_Zal}. 
	Presently, variants are classified by their physical content \cite{Raabe}. 
	Where the kinetics of phase separation must be considered, ``free--energy" methods \cite{SwiftOO96pre, m_r_swift} 
	and their thermodynamically-consistent extensions, due to Wagner et al., \cite{Wagner, Wagner_and_Li, Wagner_and_Pooley}, 
	are appropriate tools. For workers with a background in molecular simulation, the Shan--Chen method \cite{Shan_Chen} is 
	a natural choice. In continuum immiscible hydrodynamics, one incorporates dynamic conditions of stress continuity 
	(i.e. physical principles) and the kinematic condition of mutual impenetrability (with purely logical content), \cite{Landau}, as 
	boundary conditions between separate flows. In this regime it is safe to use the chromodynamic, color-gradient or phase-field method, 
	which we define as a combination of algorithms due to Lishchuk \cite{Lishchuk:02-0} (who uses earlier ideas of Brackbill, \cite{Brackbill}) and d'Ortona et al. \cite{dOrtona}.
	
	Chromodynamic MCLBE uses an immersed boundary force \cite{Brackbill, Peskin}, appropriate corrections being applied to the velocity \cite{Guo2002},
	alongside a computationally-efficient, analytic component segregation \cite{dOrtona} which distributes an interface, which, for continua, should be sharp.
	(Note, Reiss and Phillips \cite{Reiss} developed an inter-facial perturbation to replace immersed boundary forces, which is the most physically consistent encapsulation of MCLB inter-facial tension
	as a perturbation to the stress.)  The method is the most direct descendant of Gunstensen's original, in which the problems of lattice pinning and faceting have been reduced,  
	Reiss and Dellar \cite{Dellar3, Reiss0} having identified their origin and a means to reduce the impact of the unphysical interface width scale.
	Such limitations notwithstanding, chromodynamic method is robust, transparent, has low micro-current and 
	allows direct parameterization of inter-facial tension, width \cite{Halliday_PRE1_2007} and the separated fluids' viscosity contrast \cite{Xu}, 
	the interface propagation in the base model is reasonably understood \cite{Kehrwald, Subheder} (but see below)
	and different CG models have been applied successfully to numerical study of steady and unsteady flow, \cite{Liu0, Leclaire, Ba, Wen}.
	
 	Here we further investigate the fundamentals of the dynamics and kinematics of a chromodynamic MCLB interface,
	when it separates fluids at density ratio $\Lambda$.
	Our data aim to support results by Ba et al. \cite{Ba}, Wen at al. \cite{Wen} 
	who have benchmarked the technique in complex flow situations using multi-relaxation time (MRT) collision schemes and 
	generalizations of the segregation method of \cite{dOrtona}. 
	Use of a MRT collision scheme complicates the relationship between model kinematics (which originate in the 
	re-color step- see section \ref{sec_intro}) and model dynamics (which is extracted by Chapman-Enskog analysis), \cite{Burgin}.
	But MRT schemes have the decisive advantage of stability. 
	Hence, we develop a Dellar-type MRT scheme, for chromodynamic MCLBE which couples model kinematics and dynamics clearly.
	Taking this model as representative of chromodymamic MRT schemes,
	we extend previous work \cite{Burgin}, to measure the extent to which such models meet appropriate dynamic and kinematic conditions.
	To achieve this, one should consider fully transient flows. We do so, first with plane and, later, curved interfaces. By making 
	direct comparison with appended semi-analytic calculations, which invoke kinematic and dynamic conditions,
	we answer the questions to what extent do the lattice fluids move together at the interface? and 
	to what extent is the continuous traction condition met?
	We organize as follows. In Sec.~\ref{sec_intro} we present backgound detail of our model; in Sec.~\ref{sec_derivation} we 
	derive a MRT scheme for it; in Sec.~\ref{sec_results} we present and use semi-analytic tests alongside refined versions of existing tests, to assess its performance. 
	In Sec.~\ref{sec_conclusions}, we present our conclusions. Details are presented in the appendices.  
	\section{Background : Density difference chromodynamic MCLBE }
	\label{sec_intro}
	Represent red and blue fluid components by distribution functions $R_i(\mathbf{r}, t)$ and $B_i(\mathbf{r}, t)$, where:
	\begin{equation}
	f_i(\mathbf{r}, t) = R_i(\mathbf{r}, t) + B_i(\mathbf{r}, t).
	\end{equation} 
	Above, $i=0,1,..(Q-1)$ indexes the $Q$ lattice links in the model (Fig.~\ref{figA1}). 
	Let $\rho=(\rho_R+\rho_B)$, $\rho_R$, $\rho_B$, $\delta_t$, $c_{i \alpha} $, $w_i$, $u$ and $c_s$ denote nodal density, red nodal density, blue nodal density, time step, 
	the $\alpha$ component of the $i^{th}$ lattice basis vector, the weight for link $i$, fluid velocity and the color-blind speed of sound (or the geometrical lattice tensor isotropy constant). Other symbols have their usual meanings. 
	A MRT collision scheme, for a single fluid subject to a body force, $G_{\alpha}(\mathbf{r})$, has kinetic equation:
	\begin{eqnarray}
	\nonumber
	f_i(\mathbf{r}+\delta_t \mathbf{c}_i, t+\delta_t) = && f_i(\mathbf{r}, t)  - \sum_{j=0}^{Q-1} A_{ij} ( f_j(\mathbf{r}, t) - f^{(0)}_j(\rho, \mathbf{u}) )\\
	\label{equ_evolution}
	&& + F_{1 i} + F_{2 i},
	\end{eqnarray}
	where, after \cite{Ba, Wen}, equilibrium $f_i^{(0)}$ is modified to allocate mass away from rest link ($i=0$),
	generating a density contrast \cite{Ba, Wen, Liu}:
	\begin{equation}
	\label{equ_equ}
	f_i^{(0)}(\rho, \mathbf{u}) = \rho \phi_i  + w_i \rho\left( \frac{u_\alpha c_{i \alpha}}{c_s^2}+ \frac{u_\alpha u_\beta c_{i\alpha} c_{i\beta}}{2 c_s^4} - \frac{u^2}{2 c_s^2} \right),
	\end{equation}
	with:
	\begin{eqnarray}
	\label{eq_phi}
	\phi_i =
	\begin{cases}
	\frac{\alpha_R \rho_R}{\rho}+\frac{\alpha_B \rho_B}{\rho}, & i = 0, \\ \nonumber
	k w_i \left[ (1-\alpha_R)  \frac{ \rho_R}{\rho} +(1-\alpha_B)\frac{\rho_B}{\rho} \right], & i \neq 0,
	\end{cases} \\
	\end{eqnarray}
	where $k=\frac{9}{5}$, in D2Q9. Above, $\alpha_R$ and $\alpha_B$ are considered shortly when discussing
	the role of $\phi_i$.
	In Eq.~(\ref{equ_evolution}), $A_{ij}$ is a collision matrix element and ``sources" $F_{1i}$ and $F_{2i}$ correct the 
	dynamics for the effects of large density contrasts and $\mathbf{G}$ respectively \cite{Burgin}. 
	Term $F_{1i}$ is expressed in tensor Hermite polynomials:
	\begin{equation}
	\label{equ_source1}
	F_{1i} = w_i T_{\alpha \beta}(\rho_R, \rho_B, \rho^N,\Lambda, \mathbf{u}) (c_{i \alpha } c_{i \beta} - c_s^2 \delta_{\alpha \beta}),
	\end{equation}
	and to embed $\mathbf{G}$ we use the form devised by Luo \cite{Luo}: 
	\begin{eqnarray}
	\nonumber
	F_{2 i} & = & w_i \bigg(   \frac{\bf{G} \cdot \bf{c}_{i \alpha }  }{ c_s^2}  + \frac{1}{2 c_s^4 } \left( 1 - \frac{\lambda_3}{2}  \right) \times \\
	\label{equ_source2}
	&&  ( G_{\alpha} u_{\beta} + G_{\beta} u_{\alpha} ) (c_{i \alpha } c_{i \beta} - c_s^2 \delta_{\alpha \beta})\bigg).
	\end{eqnarray}
	Term $T_{\alpha \beta}$ and eigenvalue $\lambda_3$ 
	(which determines lattice fluid kinematic viscosity) are considered in Appendix~\ref{sec_appendix1}. 
	Note, we assume force-adjusted macroscopic observables: 
	\begin{eqnarray}
	\label{equ_u_def}
	(\rho_R, \rho_B) = \sum_i \left( R_i,B_i \right), \quad \mathbf{u} = \frac{\sum_i f_i(\mathbf{r}, t) \mathbf{c}_i }{\rho  } + \frac{\mathbf{G}}{2\rho}.
	\end{eqnarray}

	Return now to the density contrast mechanism embedded in $f_i^{(0)}$ and $F_{1i}$. 
	Parameters $\alpha_R$ and $\alpha_B$ are chosen such that:
	\begin{equation}
	\label{equ_fred}
	\Lambda = \frac{\rho_{0R}}{\rho_{0B}} = \frac{c_{sB}^2}{c_{sR}^2} = \left(  \frac{1-\alpha_B}{1-\alpha_R} \right),
	\end{equation}
	i.e. to control density contrast, $\Lambda$, via the sonic speed.
	Eq.~(\ref{equ_fred}) supports a condition for mechanical stability, $\rho_{0R}c^2_{R} = \rho_{0B}c^2_B$, where 
	$\rho_{0C}$ is the density deep within the component $C = R,B$. 

	Components are identified by a color index $\rho^N(\mathbf{r},t)$:
	\begin{equation}
	\label{equ_rhoN_def}
	\rho^N (\mathbf{r},t) \equiv \frac{ \left( \frac{\rho_R (\mathbf{r},t) }{\rho_{0R} } -  \frac{\rho_B (\mathbf{r},t) }{\rho_{0B} } \right) }{ \left(  \frac{\rho_R (\mathbf{r},t) }{\rho_{0R} } +  \frac{\rho_B (\mathbf{r},t) }{\rho_{0B} }  \right) } \in [-1,1],
	\end{equation}
	\cite{Ba, Wen, Liu}, in terms of which inter-facial tension is created by the action of force:
	\begin{equation}
	\label{equ_stforce}
	\mathbf{G} = \frac{1}{2} \sigma K \mathbf{\nabla} \rho^N,
	\end{equation}
	where $\sigma$ is the inter-facial tension and the mean curvature is measured as follows \cite{Brackbill}:
	\begin{equation}
	K = \underline{ \mathbf{\nabla} } \cdot \hat{n}, \quad \hat{n} = - \left( \frac{\nabla \rho^N }{ |\nabla \rho^N |} \right),
	\end{equation}
	for a red drop, with the usual convention on surface normal, $\hat{n}$. 
	Color field $\rho^N$ is considered continuous, changing rapidly only in the inter-facial region.
	Its variation may be sharpened \cite{Dellar3, Reiss0} and it may be used to control kinematic viscosity, by setting 
	$\nu(\rho^N) = \frac{1}{6} \left(\frac{2}{\lambda_3 (\rho^N)} - 1\right)$, \cite{Xu3D, Xu2D}.
	Kinetic-scale, post-collision color segregation is an adaptation of \cite{dOrtona}:
	\begin{eqnarray}
	\label{eq_re_color}
	\nonumber C_i^{++}(\mathbf{r},t) & = & \frac{\rho_C(\mathbf{r}, t)  }{\rho(\mathbf{r}, t) }f_i (\mathbf{r}, t)^+ \\ 
	& \pm & \beta \frac{\phi_i (\mathbf{r},t) \rho_R(\mathbf{r}, t) \rho_B(\mathbf{r}, t )}{\rho(\mathbf{r}, t)} \hat{\mathbf{n}} \cdot \delta _t \hat{\mathbf{c}}_i,
	\end{eqnarray}
	where superscript $+$ ($++$) denotes a post-collision (post re-color) quantity and $\beta$ is a chosen parameter \cite{dOrtona}.
	This simple segregation rule is mass-conserving, local (given a director, $\hat{\mathbf{n}}$) and 
	``bottom-up", i.e. a kinetic scale postulate. It is usually ignored in deriving macroscopic model behavior.
	However, Eq. (\ref{eq_re_color}) is consistent with a modified equation for uniform fluid motion \cite{Burgin}:
	\begin{eqnarray}
	\label{eq_rhor3}
	\nonumber && \frac{D \rho_R}{D t} +\frac{1}{2} \delta_t  \frac{\partial^2 \rho_R}{\partial t^2}  \\
	\nonumber &=& \frac{k}{2} c_s^2 (1-\alpha_R) \delta_t  \nabla^2 \left( \frac{\rho_R^2}{\rho} \right)\\
	\nonumber && + \frac{k}{2} c_s^2 (1-\alpha_B) \delta_t  \nabla^2 \left( \frac{\rho_R \rho_B}{\rho} \right)\\
	\nonumber&&  + \frac{1}{2}\delta_t  u_\alpha u_\beta \partial_\alpha \partial_\beta \rho_R \\
	\nonumber&&   -\delta_t  \beta (1-\alpha_R) k c_s^2 n_\gamma \partial_\gamma \left( \frac{\rho_R^2 \rho_B}{\rho^2} \right)   \\
	\nonumber&&  -\delta_t  \beta (1-\alpha_B) k c_s^2 n_\gamma \partial_\gamma \left( \frac{\rho_R \rho_B^2}{\rho^2} \right) \\
	&&  + 2 \delta_t  c_s^4 \partial_{\alpha} \partial_{\beta}\left( \frac{ \rho_R T_{\alpha \beta } }{ \rho} \right).
	\end{eqnarray}
	Above, the last term on the right hand side originates in correction term, $F_{1i}$ (see Eq.~(\ref{equ_source1})). 
	Burgin et al. \cite{Burgin} give this term for an LBGK collision model; on neglecting it they find by solving Eq.~(\ref{eq_rhor3}):
	$\rho_R (\mathbf{r},t)  = \frac{\rho_{0R}}{2} \big(1 + \tanh(\beta \mathbf{\hat{\mathbf{n}}} \cdot (\mathbf{r} - \mathbf{u}t) \big)$, 
	with equivalent behavior for $\rho_B$. When substituted in Eq. \ref{equ_rhoN_def}, these variations reveal a smoothly varying color index:
	\begin{eqnarray} 
	\label{equ_rhoN}
	\rho^N  (\mathbf{r},t)  = \tanh\left[ \beta \hat{\mathbf{n}} \cdot (\mathbf{r}- \mathbf{u}t ) \right].
	\end{eqnarray}
	Quantity $\rho^N$ is a material invariant, at leading order- see below.
	On the other hand, the last term in Eq.~(\ref{eq_rhor3}) constitutes an error associated with pure advection, present even in uniform flow, which is shown to restrict applicability of method.
	As remarked above, taking the order $\delta_t$ terms in Eq.~(\ref{eq_rhor3}):
	\begin{equation}
	\label{equ_kinematic_origin}
	\frac{\partial \rho_R}{\partial t} + u_\gamma \partial_\gamma \rho_R \approx 0, \quad  \frac{\partial \rho_B}{\partial t} + u_\gamma \partial_\gamma \rho_B \approx 0,
	\end{equation}
	which is useful in deriving our MRT scheme, in Sec.~(\ref{sec_derivation}), where 
	Eq.~(\ref{equ_kinematic_origin}) is taken to imply that on short timescales, $t_0$, the color index is an approximate
	material invariant, which eliminates its $t_0$ derivatives from the Euler equation.
	
	Note, Eqs.~(\ref{equ_source1}), (\ref{equ_source2}) and (\ref{equ_stforce}) require numerical gradients.
	Typically, compact second order stencils, relying on lattice isotropies are found to be sufficient in MCLBE 
	but higher order, non-compact versions (see Sec.~(\ref{sec_stencils})) are helpful, here.  
	\section{ MRT scheme for large density difference chromodynamic MCLBE}
	\label{sec_derivation}
	Dellar  \cite{Dellar2003, Dellar2008} developed an MRT scheme for single component flow, which was extended to accommodate the 
	force, $\mathbf{G}$, used in chromodynamic lattice Boltzmann multi-component flow \cite{Xu}. 
	Here, we further adapt that method to completely immiscible fluids, with density contrast $\Lambda$,
	where it is necessary to consider large density gradients in the region of rapidly changing $\rho^N$.
	
	Dellar's is arguably the most aesthetic and logically consistent MRT scheme. 
	$\mathbf{A}$ is defined by its eigenvalues and eigenvectors,
	only a subset of which must be chosen, a majority being assigned in the Chapman-Enskog process.    
	Working from a weighted orthogonal modal basis introduced by Junk \cite{Junk}, Dellar \cite{Dellar2008, Dellar2003} devised 
	a MRT scheme with less coupling between the density, momentum and stress modes and the 3 ``ghost" modes, (in D2Q9)
	than is present in the more commonly used MRT scheme of Lallemand and Luo \cite{Lallemand}.
	We derive, in Appendix~\ref{sec_appendix1}, a MRT scheme-based model, generalized to chromodynamic immiscible fluids.
	Our analysis, performed in $D2Q9$, attempts to clarify the coupling between collision and model kinematics. See also \cite{Burgin}.
	The resulting scheme involves a set of macro-scopic modes, $\mathbf{h}^{(p)}$, defined in Table~\ref{tabA1}; a majority 
	representing observables e.g. momentum components. 
	\begin{figure}[ht]
		\begin{center}
			\includegraphics[width=4cm]{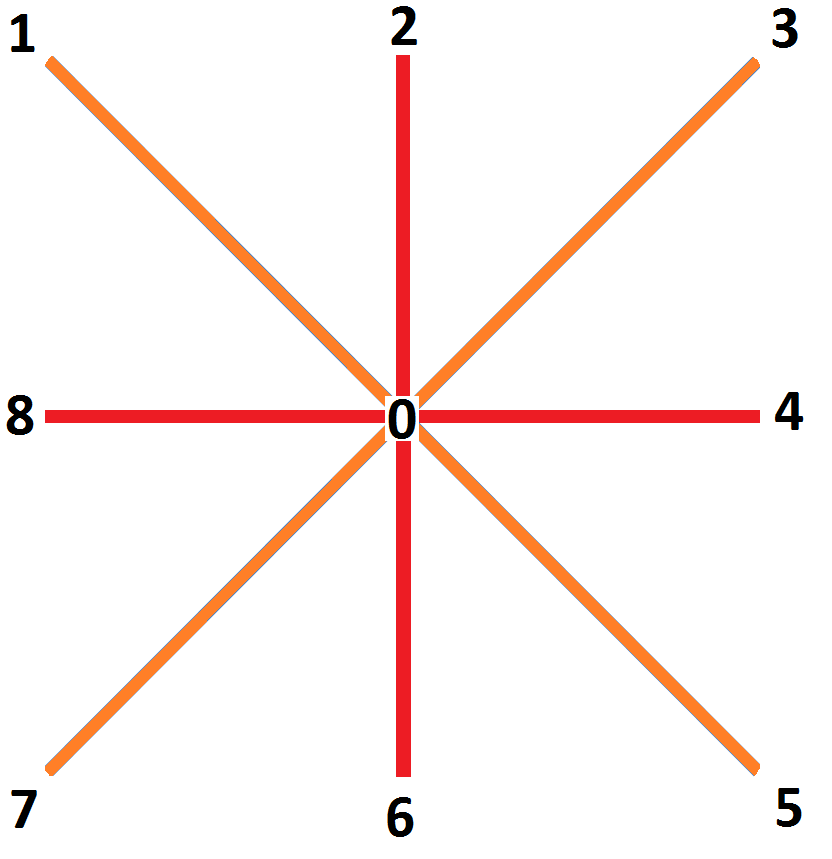}
			\caption{ Schematic. Square D2Q9 lattice with our indexing convention. Odd values of $i$ identify the longer links.}
			\label{figA1}
		\end{center}
	\end{figure}
	\begin{table*}[htb]
		\resizebox{13.7cm}{!}{
			\begin{tabular}{ |c|c|c|c|c|L|c| } 
				\hline
				eigenvector & component & definition & eigenvalue, $\lambda_p$ & mode, $m^{(p)}$ & physical interpretation & equilibrium \\ 
				\hline
				$\mathbf{h}^{(0)}$ & $h_i^{(0)}$ & $w_i$ &	0 & $\rho$ & density & $\rho$  \\ 
				\hline
				$\mathbf{h}^{(1)} $ & $h_i^{(1)}$ & $w_ic_{ix}$  &	0 & $\rho u_x$ & x momentum & $\rho u_x$  \\ 
				\hline
				$\mathbf{h}^{(2)} $ & $h_i^{(2)}$ & $w_i c_{iy}$  &	0 & $\rho u_y$ & y momentum & $\rho u_y$  \\
				\hline
				$\mathbf{h}^{(3)} $ & $h_i^{(3)}$ & $w_i c_{ix}^2$  &	$\lambda_3$ & $\Pi_{xx}$ & Momentum flux component & $\Pi_{xx}^{(0)}$  \\  
				\hline
				$\mathbf{h}^{(4)} $ & $h_i^{(4)}$ & $w_i c_{iy}^2$  &	$\lambda_3$ & $\Pi_{yy}$ & Momentum flux component & $\Pi_{yy}^{(0)}$   \\ 
				\hline
				$\mathbf{h}^{(5)} $ & $h_i^{(5)}$ & $w_i c_{ix}c_{iy}$  &	$\lambda_3$ & $\Pi_{xy}$ & Momentum flux component & $\Pi_{xy}^{(0)}$   \\ 	
				\hline
				$\mathbf{h}^{(6)} $ & $h_i^{(6)}$ & $g_i$  &	$\lambda_6$ & $N$ & - & 0 \\ 		
				\hline
				$\mathbf{h}^{(7)} $ & $h_i^{(7)}$ & $g_i c_{ix}$ &	$\lambda_7$ & $J_x$ & - & 0 \\ 
				\hline
				$\mathbf{h}^{(8)} $ & $h_i^{(8)}$ & $g_i c_{iy}$  &	$\lambda_7$ & $J_y$ & -  & 0 \\ 							
				\hline	
			\end{tabular}
		}
		\caption{ Collision matrix eigenspectrum. Left row eigenvectors (projectors), $\mathbf{h}^{(p)}, p = 0,1,...,8$, corresponding eigenvalues, corresponding physical significance (if any) and corresponding equilibria for mode $m^{(p)} \equiv \sum_i h^{(p)}_if_i$ of the collision matrix, $\mathbf{A}$.}
		\label{tabA1}	
	\end{table*}

	We define a projection matrix, comprised of orthogonal left row collision matrix eigenvectors, $\mathbf{h}^{(p)}$, each a projector of a particular mode, $m^{(p)}$, 
	\begin{equation}
	\nonumber
	\mathbf{M}\equiv \left( \mathbf{h}^{(0)}, \mathbf{h}^{(1)}, \cdots, \mathbf{h}^{(8)} \right)^T, 
	\end{equation}
	such that:
	\begin{eqnarray}
	\nonumber
	&&\left(m^{(0)},m^{(1)},...,m^{(8)}\right)^T = \mathbf{M}\ \mathbf{f} \\
	\nonumber
	&&= \left( \rho, \rho u_x, \rho u_y, \sigma_{xx}, \sigma_{yy}, \sigma_{xy}, N, J_x, J_y \right)^T,
	\end{eqnarray}
	(see Table~\ref{tabA1}). Above, column vector $\mathbf{f} \equiv \left( f_0, f_1,...,f_8 \right)^T $. We define all the $\mathbf{h}^{(p)}$ as weighted polynomial expressions in the 
	lattice basis of Fig.~\ref{figA1}, because a subset (of the $\mathbf{h}^{(p)}$) are naturally identified as such when deriving the dynamics: see Appendix~\ref{sec_appendix1}.
	Project Eq.~(\ref{equ_evolution}) using left multiplication by $\mathbf{M}$:
	\begin{equation}
	\label{equ_temp1}
	\mathbf{M \ f^+} = \mathbf{M \ f} + \mathbf{M \ A \ M^{-1}} \left( \mathbf{M \ f^{(0)}} -\mathbf{M \ f} \right) +\mathbf{M \ F},
	\end{equation}
	where $\mathbf{F}$ is the column vector whose elements are $F_i = F_{1i} + F_{2i}$. 
	The projected evolution equation decomposes to forced scalar relaxations for each mode:
	\begin{eqnarray}
	\label{equ_temp}
	\nonumber
	m^{(p)+} &=& m^{(p)} + \lambda_p \left( m^{(0) (p)} - m^{(p)} \right) + S^{(p)}, \\
	S^{(p)} &=& \sum_{j=0}^8 M_{pj} F_j, \quad p = 0,1,2,...,(Q-1).
	\end{eqnarray}
	In Eq.~(\ref{equ_temp}), we use the properties of the $\mathbf{h}^{(p)}$, from which
	$\mathbf{M\ A}=\mathbf{ \Lambda \ M} $, i.e. $\mathbf{ \Lambda} = \mathbf{M\ A \ M^{-1}}$, 
	with $\mathbf{\Lambda}\equiv diag (\lambda_0,\ \lambda_1,...,\lambda_8)$. Note, zero eigenvalues are associated with physical modes subject to conservation principles.
	Developing a MRT scheme now reduces to specifying equilibria, $ m^{(0) (p)}$, and sources $ S^{(p)}$,
	such that a Chapman-Enskog expansion of the kinetic scale dynamics predicts that the physical modes (Tab.~\ref{tabA1}) 
	conform with the continuity and Navier-Stokes equations. See Sec.~\ref{sec_appendix1}.  
	An advantage of Dellar's approach is that $\mathbf{M}$ may be inverted, using lattice isotropies. 
	The modal evolutions in Eq.~(\ref{equ_temp}) are inverted to yield $\mathbf{f^+} = \mathbf{M^{-1} \ m^+}$,
	So, post-collision distribution function is constructed directly from post-collision $m^{(p)+}$:
	\begin{eqnarray}
	\nonumber
	f_i^+&=&(M)_{ij}^{-1}\ m_j^+\\
	\nonumber
	&=& w_i \Bigg\{  \bigg[ 2-\frac{3}{2}\left( c_{ix}^2+c_{iy}^2 \right) \bigg] \rho \\
	\nonumber
	& & \ \ \ \ \ + 3 \left( (\rho u_x)^+ c_{ix} + (\rho u_y)^+ c_{iy}\right)   \\
	\nonumber
	& & \ \ \ \ \ + \frac{9}{2} \left( \Pi_{xx}^+ c_{ix}^2 +2\Pi_{xy}^+ c_{ix}c_{iy} +\Pi_{yy}^+ c_{iy}^2 \right)\\
	\nonumber
	& & \ \ \ \ \ -\frac{3}{2} \left(\Pi_{xx}^+ + \Pi_{yy}^+\right)\\
	\nonumber
	& & \ \ \ \ \ + \frac{1}{4} g_i N^+ + \frac{3}{8} g_i \left( J_x^+ c_{ix} + J_y^+ c_{iy} \right) \Bigg\},
	\end{eqnarray} 
	with $(\rho u_x)^+$, $(\rho u_y)^+$, $\rho^+$, $\Pi_{xx}^+$, $\Pi_{xy}^+$, $\Pi_{yy}^+$, $N^+$, $J_x^+$ and $J_y^+$
	given explicitly in Eqs.~(\ref{moev1} - \ref{moev7}). Of course, color is finally re-allocated according to Eq.~(\ref{eq_re_color}).
	Tensor $T_{\alpha \beta}$ in Eqs.~(\ref{equ_source1}), (\ref{eq_rhor3}) is shown, in Appendix~\ref{sec_appendix1}, Eq.~(\ref{equ_T_identity}),
	to be identical to that of Burgin et al. \cite{Burgin}, for an LBGK model.  
	\section{ Results and Discussion}
	\label{sec_results}
	The accuracy of our multi-component scheme of Sec.~\ref{sec_derivation}
	is assessed against the conditions of mutual impenetrability (model kinematics) and 
	the viscous stress transmission (model dynamics).
	Transfer of momentum between immiscible fluids is controlled by boundary conditions 
	which refer to both kinematics and dynamics. In Sec.~\ref{sec_test2} we present develop
	two transient test-bench flows which rely upon these conditions which we compare with data.
	We mainly consider, here, the dynamics of the scheme, its kinematics having been effectively assessed by Burgin et al., \cite{Burgin}, on the following argument.
	Whilst the work of Burgin et al. uses an LBGK  collision method (to highlight the connection between the model kinematics and dynamics),
	the key tests applied consider performance in uniform flow, with a flat interface i.e. $\mathbf{G} = \mathbf{0}$.
	In this regime, there is no practical distinction between the operation of MRT and LBGK schemes.
	Put another way, Burgin's simulation data applies to the chromodynamic MCLBE MRT method of Sec.~\ref{sec_derivation}.
	(Note, however, we have confirmed this explicitly).  
	Moreover, the kinetic equation source due density difference effects (see Eqs.~(\ref{equ_source1})),
	is identical to that for LBGK collision. 
	
	We consider here curved fluid-fluid interfaces, as well as plane interfaces. 
	No assessment would be complete without some assessment of the inter-facial micro-current.
	For all the data presented below, we relax the ghost modes of our MRT scheme to equilibrium i.e. $\lambda_7=\lambda_8=1$. 
	\subsection{Plane Interfaces}
\begin{figure}[H]
		\begin{center}
			\includegraphics[width=9cm]{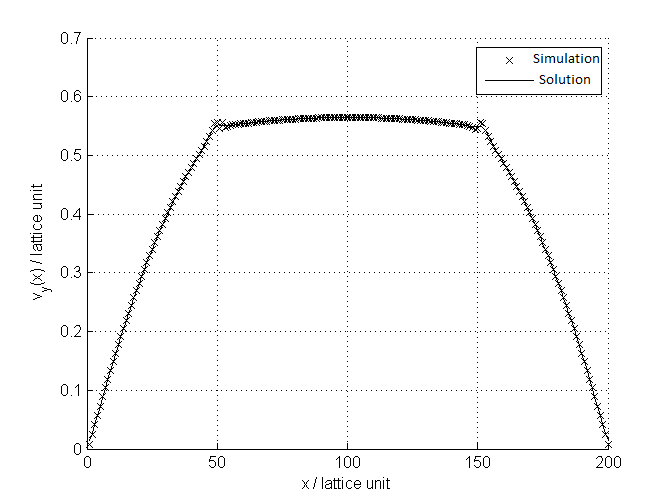}
			\caption{Transverse variation of the flow velocity for the test illustrated in Fig.~\ref{fig_BaTest}. Simulation data are represented by crosses and semi-analytic theory which accounts 
				for the transverse variation of the density is indicated by the continuous line.
				For these data, $L_x=200$, $\alpha_B = 0.2$, $\alpha_R = 0.9$, $\Lambda=8$, $\nu_1=\nu_2 = 0.333$.  }	
			\label{Ba_test}
		\end{center}
	\end{figure}

	The data in Fig.~\ref{Ba_test} compare simulation and theory. 
	We test the steady-state of uni-directional, pressure-driven flow, with the transverse density stratification illustrated in Fig.~\ref{fig_BaTest}.
	Note, we do not benchmark against the solution for discontinuous variation of density (see e.g. Ba et al., \cite{Ba}).
	Instead, we compare simulation data (crosses) with a semi-analytical solution in Appendix~\ref{sec_sol_Ba_test}, which accounts for the effects of 
	continuous variation of density at interface (continuous line). For these data, the simulation width $L_x = 200$, $\alpha_B = 0.2$, $\alpha_R = 0.9$ 
	(corresponding to a density contrast between separated components' bulk of $\Lambda = 8$) and $\nu_B = \nu_R = 0.333$.  
	These data compare well with theory and data generated by identical tests applied to the MRT schemes of Ba et al. \cite{Ba}, 
	which are based upon equivalent MCLBE interface schemes and traditional MRT collision operators. 
	Note, however, that we find it necessary to use high order stencils of Appendix~\ref{sec_stencils} to compute density gradients.   
	
	It is important to note that steady-state data in Fig.~\ref{Ba_test} do not verify instantaneous 
	compliance with kinematic (impenetrability) and dynamic (continuous traction) conditions. For that, one needs a transient flow. 
	Semi-analytical solutions for multi-component flow with flat and curved interfaces, which reference the key boundary conditions at issue 
	are derived in appendix Sec.~(\ref{sec_test2}).

\begin{figure*}[htb]
		\begin{center}
			\includegraphics[width=18cm]{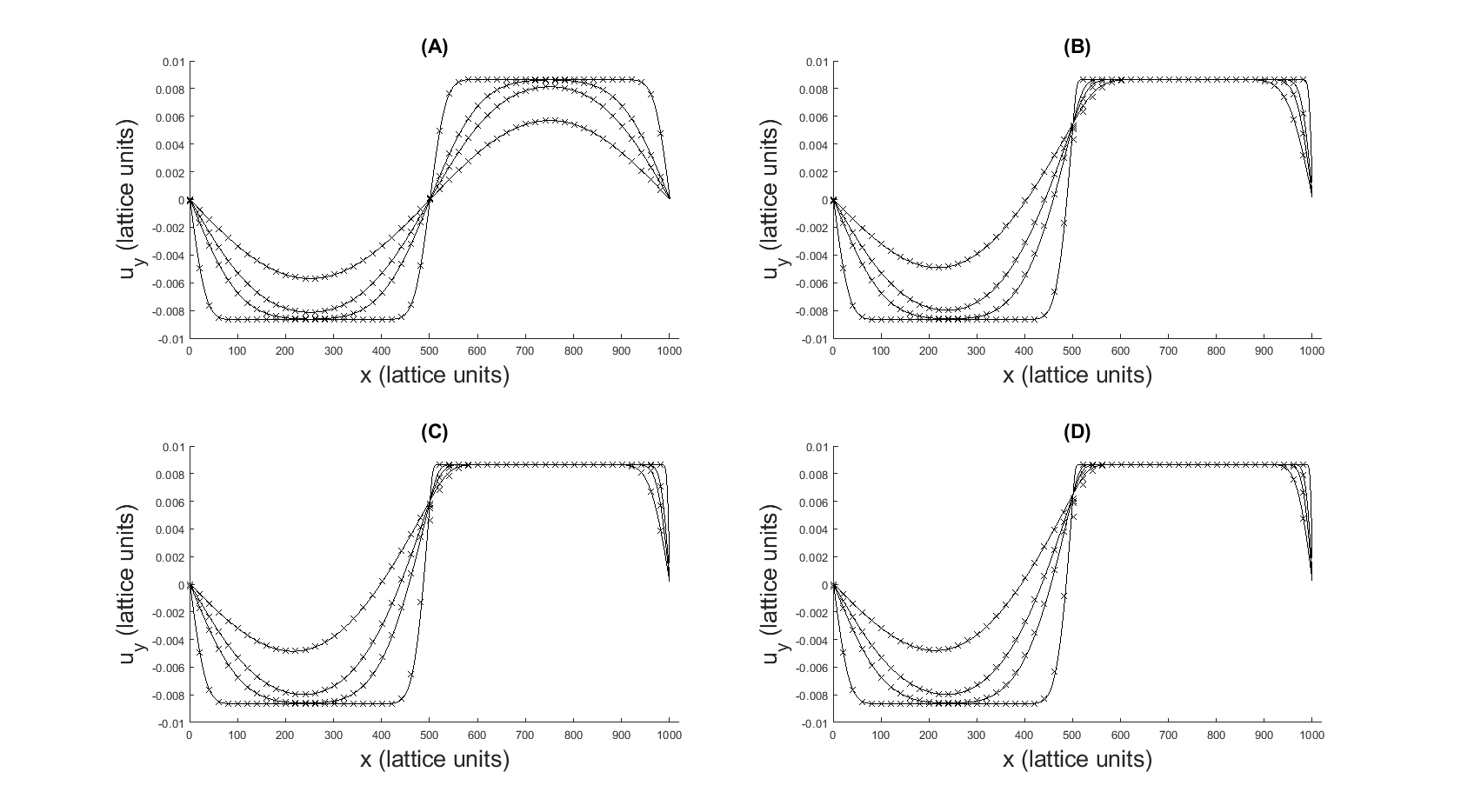}
			\caption{Comparison of simulation data (crosses) and semi-analytical solution (see Appendix~(\ref{sec_test2})) for a large range of density contrasts, $\Lambda$.
				For these data, shear viscosity $\eta =$ constant, whilst kinematic viscosity, $\nu = \frac{\eta}{\rho}$ changes.
				The interface centers on $x=500$ lattice units, with fluid on the right in all these figures is the denser fluid.
				For panels (A)..(D) $\Lambda = 1, 20, 31.25, 50$.
				These data confirm continuity of velocity and correct  transmission of stress across a flat, sheared interface. }	
			\label{fig_DynamicsTest}
		\end{center}
	\end{figure*}

	In Appendix~\ref{sec_test2}, we consider the temporal decay of a uni-directional flows of two liquids of different density separated by a 
	flat interface. The systems have defined initial velocity profile and the motion decays to rest. 
	The geometry and flow initial conditions defining our tests are shown schematically in e.g. Fig.~\ref{fig_system}.
	The density and, with it, the kinematic viscosity change at the interface, which is tangentially sheared.  
	We have obtained analytical benchmarks for this problem, in the sharp interface limit, in Appendix~\ref{sec_test2}, using Sturm-Liouville theory \cite{Arfken} straightforwardly. 
	Fig.~\ref{fig_DynamicsTest} compares simulation data (crosses) and the analytical solution, for large range of density contrasts, $\Lambda$ (see caption). 
	For these data, shear viscosity $\eta = 0.166=$ and segregation parameter $\beta = 0.5$ are constant whilst  kinematic viscosity $\nu = \frac{\eta}{\rho}$ changes. 
	This change is assumed discontinuous in Appendix~\ref{sec_test2}, whereas in simulation density varies across the interface.
	Even so, it is clear that these data confirm continuous operation of the continuous traction condition across the interface,  
	not simply that the correct \emph{steady-state} profile is obtained. This assertion is supported by the data in Tab.~(\ref{fig_DynamicsTest}), which 
	show the domain-average, relative error between the semi-analytic solution for $u(x,t)$, and the simulated solution, $u^*(x,t)$:
	\begin{equation}
	\label{equ_aps_defn}
	\epsilon (t) = \frac{ \sum_i |u(x_i,t) -u^*(x_i,t)|^2}{\max(u^*(x_i,t))^2},
	\end{equation}     
	which never exceeds $1\%$. Above, $x_i$ denotes the discrete, ``on-lattice'' value of the transverse co-ordinate.
	In Fig.~(\ref{fig_DynamicsTest}) the denser fluid is on the right.
	Its greater density means that it is not accelerated by the traction of the fluid on the left, as strongly as the
	the fluid on the left is accelerated by the traction of the fluid on the right.  

	\begin{table}[h!]
		\centering
		\resizebox{6cm}{!} 
		{%
			\begin{tabular}{| c | c | c | c | c |} 
				\hline
				\multicolumn{1}{|c|}{} & \multicolumn{4}{ c|}{Lattice relative error (\%)} \\
				\hline
				T(lu) & $\Lambda=10$ & $\Lambda=20$ & $\Lambda = 31.25$ & $\Lambda=50$ \\ [0.5ex] 
				\hline
				1000 & 0.299  & 0.560 	& 0.754  & 0.979  \\ 
				10000 & 0.114 & 0.199 	& 0.252  & 0.300  \\
				20000 & 0.080  & 0.135 	& 0.167  & 0.196   \\
				50000 & 0.047 & 0.075 	& 0.092  & 0.115 \\ [0.5ex] 
				\hline
			\end{tabular}
			
		}%
		\caption {Time variation $\epsilon (t)$, in Eq.~(\ref{equ_aps_defn}).}
		\label{tab_time_var_eps}
	\end{table}
	
	Note that data were matched between simulation and theory by equating the non-dimensional groups 
	which scale the MCLBE dynamics and the corresponding unidirectional Navier-Stokes equation (Eq.~\ref{eq:Main}), as follows :
	$\frac{\nu(\lambda_3)^* T^*}{ H^{* 2}} = \frac{ \nu T }{ H^2 }$, 
	where the quantities with (without) asterisks are in lattice (physical) units. From this, we find the 
	simulation time corresponding to physical time $T$ as:
	\begin{equation}
	T^* =\frac{ \nu  }{\nu(\lambda_3)} \left( \frac{H^*}{H^2 } \right)^2  T 
	\end{equation}
	\subsection{Curved Interfaces}
	Consider now curved interfaces in two dimensions. 
	The expected dependence of the inter-facial pressure step on surface tension parameter, $\sigma$ was, naturally, confirmed for the range of 
	$\Lambda \in[10^{-3},10^3]$ (the range of data in Tables~\ref{tab_micro1} and \ref{tab_micro2}) and $\sigma \in[0,0.2]$.
	We proceed to consider other tests. 
	\subsubsection{Inter-facial Micro-current}
\begin{figure}[ht]
		\begin{center}
			\includegraphics[width=6cm]{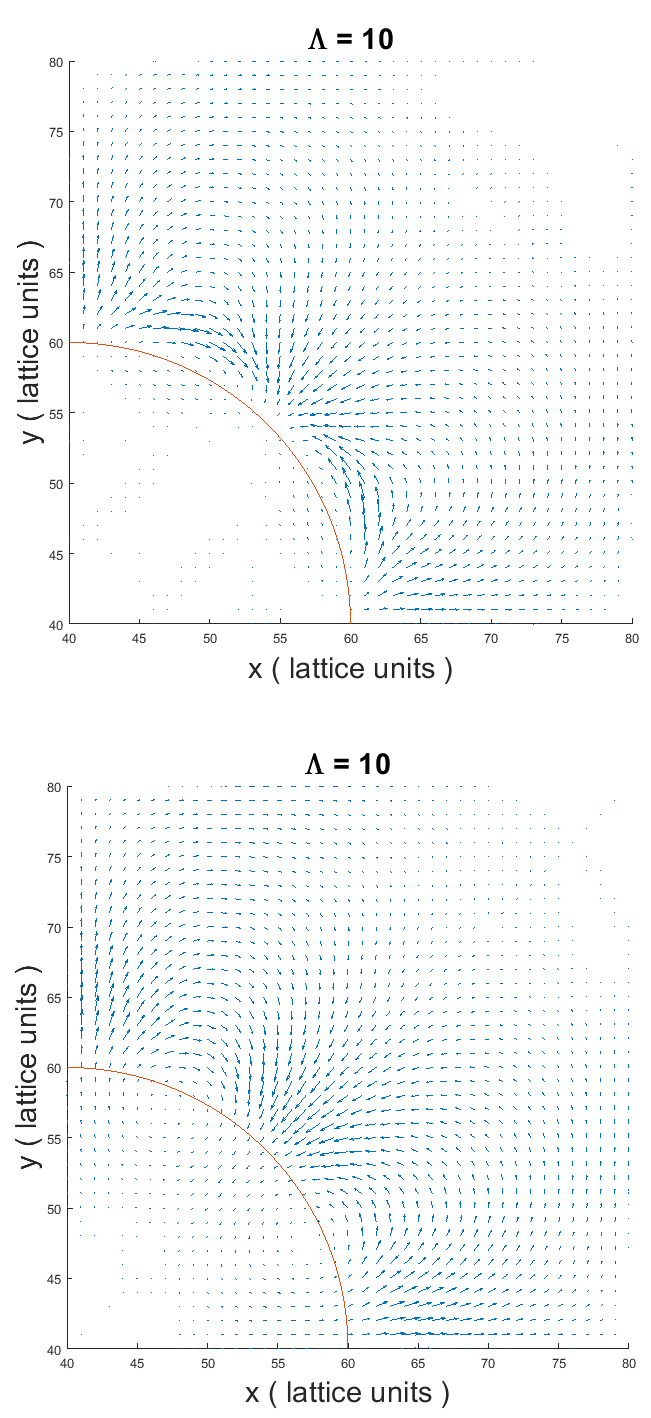}
			\caption{ Normalized micro-current flow excerpt for $\Lambda = 10$ in the vicinity of a drop, radius $R=60$ for fixed  
				$K=\frac{1}{R}$ (left) and numerically calculated curvature $K = \mathbf{\nabla}_s \rho^N$(right).
				See Tables~\ref{tab_micro1} and \ref{tab_micro2} to scale these velocity fields. The circulation in the case of fixed curvature is more localized.  
			}
			\label{fig_micro_flow}
		\end{center}
	\end{figure}

	We study a red drop, initialized with radius $R=60$, on a lattice of size $200\times200$, with periodic boundary conditions.  
	An inter-facial micro-current is present in all MCLBE models- see Fig.~\ref{fig_micro_flow}.
	It has been argued \cite{Halliday_micro} that micro-current circulation is a ``correct'' hydrodynamic response to application of a force, or perturbation, 
	which is not native to the continuum scale (where an interface is discontinuous).
	It might be argued that a micro-current is a correct hydrodynamic response to an incorrect external force. We return to this point shortly.
	For the particular case of chromodynamic MCLBE, the spatial pattern of non-isotropic numerical errors not offset by pressure (density) changes drive a persistent circulation. 
	The source of numerical error lies in derivatives, discretization error associated with the Chapman-Enskog and the re-color step.
	With an interface force, setting $K=\frac{1}{R}$ (i.e. circumventing a numerical calculation of $K$) after Eq.~(\ref{equ_stforce}) significantly reduces micro-current activity \cite{Halliday_micro}. 
	Figure \ref{fig_micro_flow} below compares the micro-current flow field, at $\Lambda = 10$, for calculated and fixed curvature drops. 
	Flow field vectors are normalized in each plot. 
	The flow in the case of fixed curvature is actually much weaker (refer to Tables~\ref{tab_micro1} and \ref{tab_micro2}) and more restricted to the inter-facial region. 
	We will return to this matter shortly.
	
	With $\Lambda=1$ (no density contrast), numerical error derives only from the interface force, with the dominant contribution arising from calculation of local interface curvature, $K$.
	In the presence of component density differences, we introduce a need to correct the dynamics, which, as we see in Sec.~(\ref{sec_derivation}), introduces strong inter-facial density gradients.
	Evolution equation source terms which rely on numerical derivatives of density add error to that already present 
	in the Lishchuk, or interface, force. Here, we make a quantitative assessment of the impact of that additional error.
	We present micro-current data for a range of separated components' density contrast, $\Lambda$, in Tables~\ref{tab_micro1} (fixed $K$) and \ref{tab_micro2}.
	Based on the above discussion, the magnitude of the micro-current depends $\Lambda$ (and, of course, $|\mathbf{G}|$),
	but is largely independent of collision scheme. This is confirmed in the data in Tables~\ref{tab_micro1} and \ref{tab_micro2}. 
	(We note that changing the collision model to an LBGK scheme does not alter any of these data by more that a few percent.) 
	For small $\Lambda$, when density contrast correction terms are small, the domain maximum micro-current flow velocity magnitude, $|\mathbf{u}|_{max} = \max( |\mathbf{u}|)$, is small.
	As the value of $\Lambda$ increases (or decreases, in case of a rare drop) the micro-current intensity increases.
	For small $\Lambda$, the micro-current regime is different, now being dominated by the interface force. 
	First, we note a dramatic reduction in micro-current recorded in both Tables~\ref{tab_micro1} (fixed $K$) and \ref{tab_micro2}, as inter-facial density gradients reduce in size. 
	Second, in comparing data for $\Lambda\in [10,0.1]$ \emph{between} Tables~\ref{tab_micro1} (fixed $K$) and \ref{tab_micro2}, we observe the signature 
	reduction in micro-current activity when we eliminate reliance on a $K$ computed from second numerical gradients.
	For larger density contrasts, where the principal cause of the circulation is presumably density contrast, the data of Tables~\ref{tab_micro1} and \ref{tab_micro2} both 
	comply with a scaling $|\mathbf{u_{max}}| \sim 7.4 \times 10^{-3} \Lambda$.
	\begin{table}[h!]
		\centering
		\begin{tabular}{||c c c c||} 
			\hline
			\multicolumn{4}{|c|}{MRT : Fixed K} \\
			\hline
			\hline
			\(\Lambda\) & \(\alpha_B\) & \(\alpha_R \) & \( |\textbf{u}|_{max} \times 10^5 \) \\ [0.5ex] 
			\hline\hline
			0.001 & 0.9995 & 0.5000 & 11.2 \\ 
			0.010  & 0.9950  & 0.5000 & 1.0 \\
			0.100   & 0.9500   & 0.5000 & 3.64\(\times 10^{-4}\) \\
			10    & 0.5000    & 0.9500 & 1.32\(\times 10^{-4}\) \\
			100   & 0.5000    & 0.9950 & 1.8 \\
			1000  & 0.5000    & 0.9995 & 7.8 \\ [1ex] 
			\hline
		\end{tabular}
		\caption{Micro-current activity for a range of separated components' density contrast.
			For these data, the interface curvature calculation (see  Eq.~(\ref{equ_stforce})) has been replaced by assigning $K=\frac{1}{R}$.
			The full flow field for the case of $\Lambda = 10$ is shown in Fig.~\ref{fig_micro_flow} (top). }
		\label{tab_micro1}
	\end{table}
	\begin{table}[h!]
		\centering
		\begin{tabular}{||c c c c||} 
			\hline
			\multicolumn{4}{|c|}{MRT : Calc K} \\
			\hline
			\hline
			\(\Lambda\) & \(\alpha_B\) & \(\alpha_R \) & \( |\textbf{u}|_{max} \times 10^5 \) \\ [0.5ex] 
			\hline\hline
			0.001 & 0.9995 & 0.5000    & 11.2 \\ 
			0.010  & 0.9950  & 0.5000    & 3.3 \\
			0.100   & 0.9500   & 0.5000    & 1.0 \(\times 10^{-1}\) \\
			10    & 0.5000    & 0.9500  & 1.8 \(\times 10^{-2}\)\\
			100   & 0.5000    & 0.9950  & 1.8 \\
			1000  & 0.5000    & 0.9995 & 7.8 \\ [1ex] 
			\hline
		\end{tabular}
		\caption{Micro-current activity for a range of separated components' density contrast.
			The full flow field for the case of $\Lambda = 10$ is shown in Fig.~\ref{fig_micro_flow} (bottom). }
		\label{tab_micro2}
	\end{table}
	\subsubsection{Kinematics of Curved Interfaces}
\begin{figure*}[htb]
		\begin{center}
			\includegraphics[width=18cm]{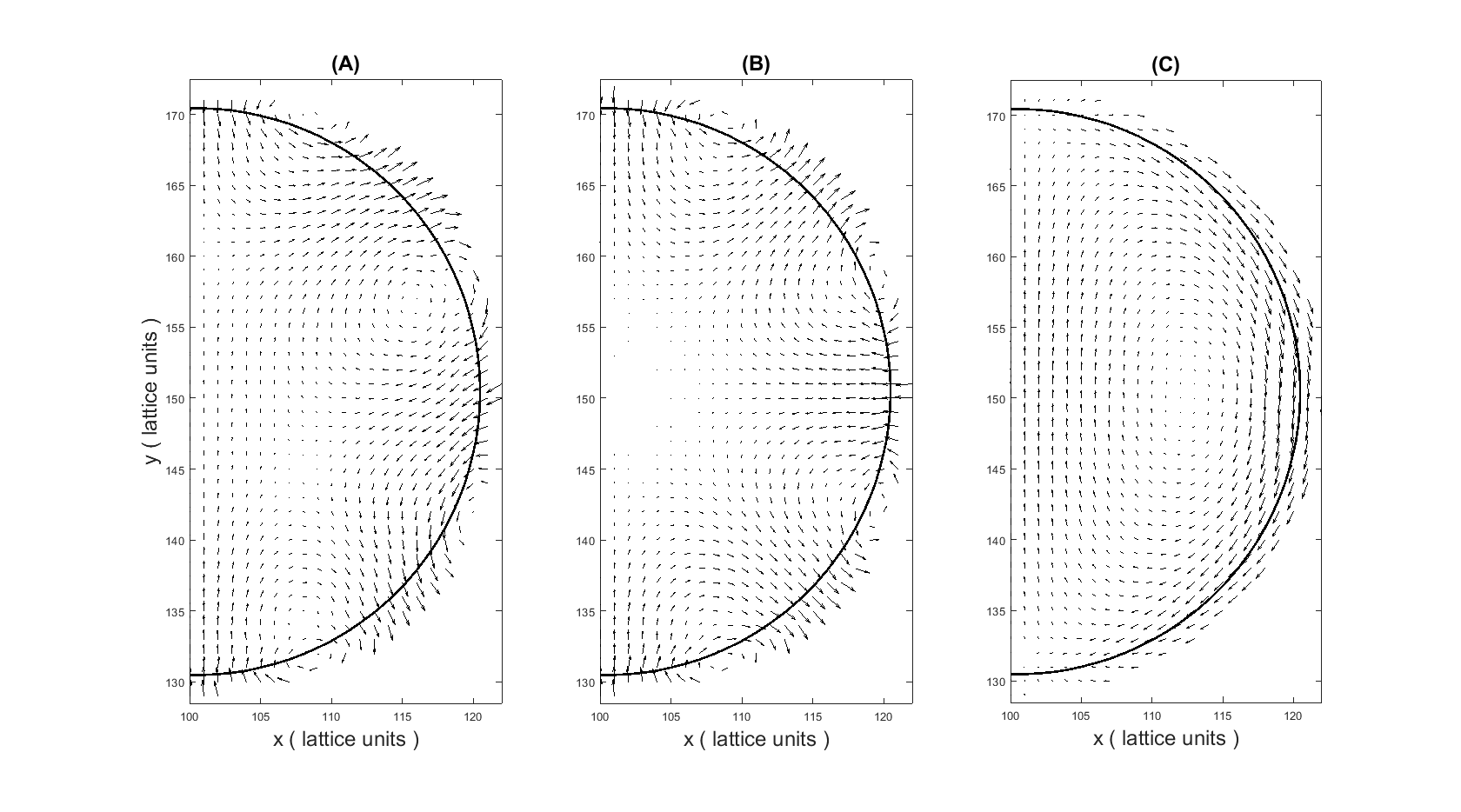}
			\caption{Low Re internal flow past a tethered, cylindrical drop for $\Lambda = 5$. 
				Flow in outside the drop has been suppressed. 
				Panel (A) shows the total flow, in which the velocity field clearly has a 
				non-physical component perpendicular to the interface. 
				(B) shows the micro-current error, measured from the frozen phase field in (A), without external flow and (C) shows the physical flow exposed by subtracting the microcurrent.
				The solid black line represents the centre of the interface between the fluids ($\rho^N = 0$ contour). The internal and external flows are clearly parallel to the interface .}	
			\label{fig_drop_flow}
		\end{center}
	\end{figure*}

	Previous work \cite{Burgin} considered kinematics of a flat interface. 
	Fig.~\ref{fig_drop_flow} (C) shows the flow (once the micro-current is subtracted), 
	which is produced when blue fluid passes a tethered, cylindrical red drop, for density contrast $\Lambda = 5$. 
	The Reynolds number must be kept very small here, to restrict deformation, and the drop is held spherical by large surface tension.
	Hence these data correspond to the challenging regime of small Reynolds and capillary number. 
	This accounts for the large micro-current. The resulting Stokes' regime flow of internal and external fluid 
	is apparently tangential to the curved interface at all points and continuous across it i.e. we observe 
	that, in the inter-facial region, $v_n = 0 $, $v_t =$ continuous.
	This accords with the kinematic condition of mutual impenetrability. 
	Note that the flow in Fig.~\ref{fig_drop_flow} (C) is not the solved flow 
	past a three-dimensional spherical drop.
	\subsubsection{Dynamics of Curved Interfaces}
\begin{figure*}[htb]
		\begin{center}
			\includegraphics[width=16cm]{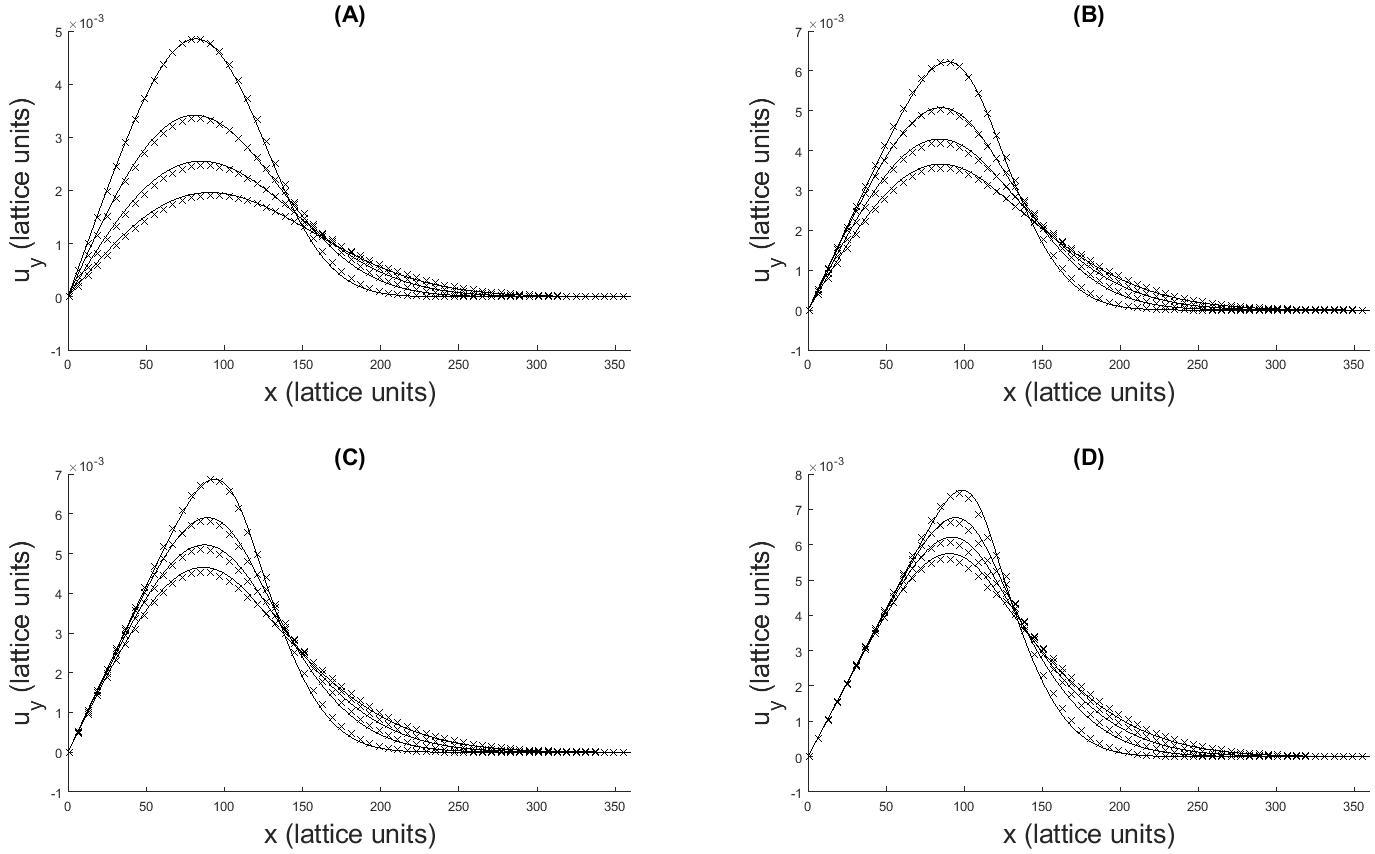}
			\caption{ Comparison of simulation data (crosses) and semi-analytical solution (see Sec.~(\ref{sec_test2})) for a range of density contrasts, $\Lambda$.
				For these data, shear viscosity $\eta =$ constant, whilst kinematic viscosity, $\nu = \frac{\eta}{\rho}$ changes.
				In all these figures, the interface centers on $r=120$ lattice units, with fluid on the left  the denser.
				For panels (A)..(D) $\Lambda = 1, 2, 3, 5$, which is smaller than the range of $\Lambda$ shown in Fig.~(\ref{fig_DynamicsTest}), note. 
				These data confirm continuity of velocity and correct  transmission of stress across a curved interface.}
			\label{fig_DynamicsTest2}
		\end{center}
	\end{figure*}

	In appendix Sec.~(\ref{sec_test2}), we consider the temporal decay of a ``unidirectional" flow of two liquids of different density separated by a curved interface.
	For this test, the system again has a defined initial velocity profile and the motion decays to rest. 
	The geometry and flow initial conditions defining our test are shown schematically in Fig.~\ref{fig_system}.
	The assumed density and, with it, the kinematic viscosity change at the interface, which is tangentially sheared. 
	In all cases, the denser fluid is on the left, which accounts for its smaller acceleration. 
	We have obtained an analytical solution for this problem, in the sharp interface limit in appendix Sec.~(\ref{sec_test2}), using adapted Sturm-Liouville theory. 
	Fig.~\ref{fig_DynamicsTest2} compares simulation data (crosses) and the analytical solution, for range of density contrasts, $\Lambda$ (see caption) which is, note,
	smaller that that in Fig.~\ref{fig_DynamicsTest2}. This reduction reflects the introduction of a curved interface. 
	For these data, $R_0 = 120$, $R=360$, shear viscosity $\eta = 0.333 =$ segregation parameter $\beta = 0.3$ are constant whilst kinematic viscosity $\nu = \frac{\eta}{\rho}$ changes. 
	This change is assumed discontinuous in the treatment of appended Sec.~(\ref{sec_test2}), whereas in simulation density varies across the interface.
	Even so, these data confirm correct transient transmission of stress across the interface in our model, not simply that the correct \emph{steady-state} profile is obtained.

	Introduction of curvature undoubtedly reduces range of density contrast available to method
	but get correct inter-facial conditions but, in general, data presented in this section confirm that 
	chromodynamic MRT schemes with density difference do recover correct boundary conditions at interface. 
	\section{Conclusions}
	\label{sec_conclusions}
	Using a single fluid formulation, we have developed a convenient, multiple-relaxation time (MRT) collision scheme multi-component
	lattice Boltzmann scheme (MCLBE) for simulating completely immiscible fluids with a density contrast, $\Lambda$, using the chromodynamic variant. 
	Our technique is based upon the method of Dellar \cite{Dellar2003, Dellar2008}.
	The model evolves a set of physical and non-physical (ghost) modes of the system, equal in number to the cardinality of lattice basis set, then 
	constructs an explicit distribution function \emph{a posteriori}.
	We place all corrections to the target dynamics (the weakly compressible Navier-Stokes equations) in the kinetic-scale evolution equation. 
	Significantly, the latter rely on density gradients, which can be large when $\Lambda$ is large, which limits applications to moderate density contrast.
	We present in the appendices enhanced (but non-compact) stencils for gradient calculation which improve performance. 
	
	Equivalent MRT schemes, due to Ba et al. \cite{Ba} and, earlier, Liu et al. \cite{Liu} pioneered our essential approach.
	These authors showed the clear benefits of MRT collision models in benchmarking against complex flow simulations. 
	To compliment this work, we focus, here, on fundamental, physical compliance in chromodynamic MCLBE MRT schemes. 
	We produce data which compare well with the steady-state tests devised by Ba et al. \cite{Ba}, but also with new theory, as follows.
	We assess our model dynamics against a semi-analytical solutions to a transient flow test cases which 
	reference, explicitly, the kinematic condition of mutual impenetrability and dynamic interface boundary condition of continuous traction. 
	Broadly, data compare well with these solutions, confirming satisfactory, instantaneous compliance with kinematic and dynamic conditions at the simulation interface. 
	Whilst the Dellar-type MRT scheme we develop here is operationally equivalent to that of Ba et al., it has an advantage.
	Practically, it has improved implementability- post collision distribution function is explicitly constructed from modes 
	with simple, scalar relaxation. Theoretically, the connection between model kinematics and dynamics is visible.
	This is a consequence of placing all density-difference dynamics corrections in the kinetic scale source term.
	MCLBE MRT schemes are not without limitations. The well-known MCLBE inter-facial micro-current. Here our simulations of curved interfaces suggest that it it may be removed completely
	from steady state simulations. Further, data presented for curved interfaces conform to our understanding of the inter-facial micro-current 
	(see \cite{Halliday_micro}) but the expected effect of dynamics corrective terms increases micro-current activity associated with the method, roughly in proportion to 
	$\Lambda$, with the contribution to the spurious signal greater than that arising from the surface tension perturbation for $\Lambda > 10$. 
	
	%
	%
	%
	\pagebreak
	
	\appendix
	
	\begin{Large}
		\bf{Appendices}
	\end{Large}
	\section{Multi-relaxation-time scheme for forced, diphasic fluids with large density contrasts }
	\label{sec_appendix1}
	\vspace{5 mm}
	We derive the Navier-Stokes equations from the multiple-relaxation-time (MRT) lattice Boltzmann equation, adapted for multi-component applications with a
	large density difference between completely immiscible components, where a body force is present. The latter is necessary to carry the interface force.   
	
	In the interest of a compact literature, we retain the overall structure of the analyses of Guo et. al. \cite{Guo2002}, 
	Dellar \cite{Dellar2008, Dellar2003} and Hou et. al.\cite{Hou1995}. Our analysis, whilst based in $D2Q9$, generalizes straightforwardly. 
	We choose to extend the scheme of Dellar because it is efficient (due to a careful choice of non-hydrodynamic modes $N$, $J_x$ and $J_y$), 
	robust, straightforward to implement and, not least, logical. In this section $\sum_i$ is used as an abbreviation for $\sum_i^{(Q-1)}$.
	
	At the kinetic scale, the forced MRT LBE  for a system subject to an ``external'' force term can be expressed as:
	\begin{eqnarray}
	\label{evo}
	\nonumber
	&& f_i\left(\mathbf{x}+\mathbf{c}_i\delta_t,t+\delta_t\right)=f_i(\mathbf{x},t)\\
	&&\ \ \ \ \ \ \ \ \ \ +\sum _{j} A_{ij} \left[f_j^{(0)}(\mathbf{x},t)-f_j(\mathbf{x},t) \right]+\delta_t F_i,
	\end{eqnarray}
	where the density-difference supporting equilibrium which distributes mass away from the rest $(j=0)$ link via term $\phi_j$, is in the form of:
	\begin{equation}
	f_j^{(0)}= \rho \phi_j +  \rho w_j \left(3u_\alpha c_{j\alpha} + \frac{9}{2} u_\alpha u_\beta c_{j\alpha} c_{j\beta}- \frac{3}{2} u_\gamma u_\gamma \right),
	\end{equation}
	and where the kinetic equation source term, $F_i$, is assumed to have the following properties:
	\begin{equation}
	\label{equ_Fis}
	\sum_{i} \left(1, \mathbf{c}_i,  \mathbf{c}_i \mathbf{c}_i \right) F_i =  \left( 0, n \mathbf{G}, \mathbf{C} + \mathbf{C}^T \right),
	\end{equation}
	where scalar $n$ and symmetric tensor, $\left( \mathbf{C} + \mathbf{C}^T \right)$, are to be determined.
	We first set-out the basics, then proceed to the Chapman-Enskog analysis to obtain the thermodynamic limit of the kinetic scheme 
	defined in Eq.~(\ref{evo}) (i.e. find appropriate expressions for tensor $\mathbf{C}$, which represents the crux of the problem of recovering correct hydrodynamics with the MRT scheme), 
	then we transform to a modal description, and finally, we invert that transformation to obtain an explicit expression for the post collision distribution function. 
	To maintain parity with the analysis of Guo et. al. \cite{Guo2002} at the outset, we now relax the definition of lattice velocity in Eq.~(\ref{equ_u_def}) 
	as follows:
	\begin{equation}
	\label{equ_bert2}
	\rho \mathbf{u} = \sum_i f_i(\mathbf{r}, t) \mathbf{c}_i  + m \mathbf{G},
	\end{equation}
	with $m$ a constant to be determined.
	
	Dellar's \cite{Dellar2008, Dellar2003} eigenvalues and corresponding left row eigenvectors for the collision matrix $A_{ij}$ can be tabulated as in Table~\ref{tabA1}, where we define:
	\begin{equation}
	\label{equ_flux_stress}
	\Pi_{\alpha\beta} \equiv \Pi_{\alpha\beta}^{(0)} +\Pi_{\alpha\beta}^{(1)},
	\end{equation}
	for $\alpha, \ \beta = x,y $, and the $\Pi_{\alpha\beta}^{(p)}$ have the usual meaning:
	\begin{equation}
	\label{equ_Pis}
	\Pi_{\alpha \beta}^{(p)} = \sum_i f_i^{(p)} c_{i \alpha} c_{i \beta}, \quad p = 0, 1. 
	\end{equation}
	Mode $\Pi_{\alpha\beta}$ will be seen, shortly, to include the momentum flux and viscous stress tensors.  
	As set-out in Table~\ref{tabA1}, matrix $A_{ij}$ has the following properties which, it will be seen, are necessary to recover correct hydrodynamics:
	\begin{eqnarray}
	\label{Apro1}\sum_i (1_{i},  c_{i\alpha}, c_{i\alpha}c_{i\beta} ) A_{ij}&=& ( 0, 0 ,  \lambda_3 c_{j\alpha} c_{j\beta}).
	\end{eqnarray}
	Here $\alpha$ and $\beta$ represent either $x$ or $y$. We also assume that the lattice basis $\mathbf{c}_i$ and the corresponding weights $w_i$ have properties:
	\begin{eqnarray}
	\label{equ_moments}
	\nonumber \sum_i w_i &=& 1,\\
	\nonumber \sum_i w_i (c_{i\alpha}) ^{2p+1}&=& 0, \ \ \ p\geq 0 \\
	\nonumber \sum_i w_i c_{i\alpha} c_{i\beta}&=& \frac{1}{3} \delta _{\alpha\beta}, \\
	\nonumber \sum_i w_i c_{i\alpha} c_{i\beta} c_{i\gamma} c_{i\theta}&=& \frac{1}{9} \left(\delta _{\alpha\beta}\delta _{\gamma\theta} +\delta _{\alpha\gamma}\delta _{\beta\theta}+\delta _{\alpha\theta}\delta _{\beta\gamma}\right),\\ 
	\end{eqnarray}
	where $\delta_{\alpha \beta}$ is the Kronecker delta. 
	Weightings $w_i$ are those of Qian et. al. \cite{Qian} and, later, Hou et. al. \cite{Hou1995}: $w_0=\frac{4}{9}$, $w_{odd}=\frac{1}{36}$, $w_{even}=\frac{1}{9}$. Fig.~\ref{figA1} 
	shows our definition and indexing of links. 
	Note that the six left row eigenvectors $\mathbf{h}^{(0)} \cdots \mathbf{h}^{(5)}$ which appear in Eqs.~(\ref{Apro1}) and Table~\ref{tabA1},	
	are linearly independent but not orthogonal. We will return to this matter. We follow Dellar \cite{Dellar2008, Dellar2003} in selecting the other
	three ``ghost'' eigenvectors, or basis vectors (see Table~\ref{tabA1}) as:
	\begin{equation}
	g_0 = 1, \quad g_{odd} = 4, \quad g_{even} = -2.
	\end{equation}
	We note that Benzi et. al \cite{Benzi1990, Benzi1992} used a qualitatively similar basis. 
	
	Our equilibrium distribution function $f_i^{(0)}$ may easily be shown to have the following necessary properties:
	\begin{eqnarray}
	\nonumber
	& & \sum_{i} \left[ 1,c_{i\alpha}, c_{i\alpha} c_{i\beta} \right] f_{i}^{(0)}(\rho,{\bf u}) = \big[ \rho,  \rho u_{\alpha} ,  \\
	& & ( 2 \phi_{1} + 4 \phi_{2}) \rho \delta_{\alpha \beta} + \rho u_{\alpha} u_{\beta} \big].
	\label{eq_appb1}
	\end{eqnarray}
	Note, $\phi_1$ and $\phi_2$ depend upon the chromodynamic field (see Eq.~(\ref{eq_phi})), so the spatial-temporal 
	variation of the isotropic term of the second moment is modified: $\sum_{i} f_{i}^{(0)} c_{i\alpha} c_{i\beta} = \left[ \frac{3}{5} \left( (1-\alpha_R) \rho_R + (1 - \alpha_B ) \rho_B \right) \delta_{\alpha \beta} + \rho u_{\alpha} u_{\beta} \right]$,
	with the variation of the speed of sound between blue components now apparent, since $c_{sR}^2 = \frac{3}{5}(1 - \alpha_R)$, $c_{sB}^2=\frac{3}{5} (1 - \alpha_B)$. 
	
	We now proceed with a Chapman-Enskog expansion of the kinetic equation and distribution function. 
	To reflect the changes occurring at different time scales, write:
	\begin{equation}
	f_i=f_i^{(0)}+\epsilon f_i^{(1)}+\epsilon^2 f_i^{(2)}+\cdots,
	\end{equation}
	\begin{equation}
	\frac{\partial}{\partial t} = \frac{\partial}{\partial t_0}+\epsilon \frac{\partial}{\partial t_1}+ \epsilon^2 \frac{\partial}{\partial t_2}\cdots
	\end{equation}
	Parameter $\epsilon$ can be interpreted as the Knudsen number.
	Assuming that density and velocity are not to be expanded in $\epsilon$, 
	the assumptions in Eq.~(\ref{eq_appb1}) imply $\sum_i f_i^{(p)}=0$ and $\sum_i f_i^{(p+1)} \mathbf{c}_i=0$, $(p\geq 1)$, but note 
	Eq.~(\ref{equ_bert2}) implies $\sum_i f_i^{(1)} \mathbf{c}_i=-m\mathbf{G}\delta_t$. 
	
	Consider the most rapid behavior in the model. Applying the above expansions, we have:
	\begin{equation}
	\label{oe} 
	O(\epsilon): \ \ (c_{i\alpha}\partial_\alpha +\partial_{t_0} ) f_i^{(0)} =-\frac{1}{\delta_t}\sum_{j} A_{ij}f_j^{(1)} +F_i .
	\end{equation}
	Summing (i.e $\sum_i$) Eq.~(\ref{oe}) and using Eqs.~(\ref{equ_Fis}), (\ref{Apro1}):
	\begin{equation}
	\frac{D}{Dt_0}\rho=0.
	\end{equation}
	For the counterpart result in the model kinematics, we use Eq.~(\ref{equ_kinematic_origin}) as $\frac{D \rho_R}{Dt} = \frac{D \rho_B}{Dt}=0$, 
	from which \cite{Burgin}:
	\begin{equation}
	\label{equ_advection}
	\frac{D \rho^N}{Dt_0} = 0.
	\end{equation}
	Multiplying Eq.~(\ref{oe}) by $c_{ix}$ (say), summing and using  Eqs.~(\ref{equ_Fis}), (\ref{Apro1}) results in an Euler equation:
	\begin{equation}
	\label{10beq}
	\partial_\alpha \Pi_{\alpha x}^{(0)} +\partial_{t_0} \rho u_x= n G_x,
	\end{equation}
	Eq.~(\ref{10beq}) differes from (10b) in \cite{Guo2002}, since the latter couples $n$, $m$ and $\tau$ (LBGK collision parameter).
	Here, we recover the appropriate Euler equation by setting $n=1$ with, note, no constraint on $m$ at $O(\epsilon)$.

	At slower $O(\epsilon)^2$, the Chapman-Enskog expansion is:
	\begin{eqnarray}
	\label{oe2}
	\nonumber O(\epsilon^2): \ \  & &\partial_{t_1} f_i^{(0)}+ (c_{i\alpha} \partial_\alpha +\partial_{t_0}) f_i^{(1)}\\
	\nonumber & &- \frac{1}{2} \left(c_{i\alpha} \partial_\alpha +\partial_{t_0}\right) \sum_j A_{ij} f_j^{(1)}\\
	= & & -\frac{1}{2}(c_{i\alpha} \partial_\alpha +\partial_{t_0}) \delta_t F_i.
	\end{eqnarray}
	Summing (i.e. $\sum_i$) Eq.~(\ref{oe2}) and simplifying gives:
	\begin{equation}
	\partial_{t_1} \rho = 0, 
	\end{equation}
	having set $\left(m-\frac{n}{2}\right) = 0$. This is equivalent to (13a) of \cite{Guo2002}. 
	Constants $m$ and $n$ are now determined for our MRT scheme:
	\begin{equation} 
	\label{equ_constants}
	n=1, \quad m=\frac{1}{2}.
	\end{equation}  
	Multiply Eq.~(\ref{oe2}) by $c_{iy}$ (say), sum, identify the
	second order moment using Eq.~(\ref{equ_Fis} ) and use Eqs.~(\ref{Apro1}), (\ref{equ_constants}):
	\begin{equation}
	\label{13beq}
	\partial_{t_1} (\rho u_y) = \partial_\alpha \sigma'_{\alpha y},
	\end{equation}
	where the viscous stress tensor $\sigma'_{\alpha y}$ is:
	\begin{equation}
	\label{equ_viscous_stress}
	\sigma'_{\alpha y} = -\left(1-\frac{\lambda_3}{2}\right) \Pi_{\alpha y}^{(1)} - \frac{\delta_t}{4} (C_{\alpha y}+C_{y\alpha}).
	\end{equation}
	Eq.~(\ref{13beq}) is the MRT equivalent of Eq.~(13b) in \cite{Guo2002}. (Our assignment $m=\frac{1}{2}$ accords with Guo et. al. 
	but their constraint $\left(n+\frac{m}{\tau}\right)=1$ does not arise here).
	
	So far, our approach has parallels that of \cite{Guo2002} but our use of an MRT scheme means we must proceed to an expression for $\Pi_{\alpha \beta}^{(1)}$ via a 
	second moment of Eq.~(\ref{oe}) (i.e. multiply by $c_{i \alpha} c_{i \beta}$ and sum). After algebra:
	\begin{equation}
	\begin{split}
	\label{Pi_1}
	\lambda_3 \frac{\Pi_{\alpha \beta}^{(1)}}{\delta_t} =& - \frac{2 \rho}{3} S_{\alpha \beta} - u_\alpha \left(G_\beta - \partial_\beta \Phi' \right)  -u_\beta \left(G_\alpha - \partial_\alpha \Phi' \right) 
	\\
	& + \left[ u_\gamma \partial_\gamma \Phi' - \frac{1}{3} \rho \partial_\gamma u_\gamma \right]\delta_{\alpha \beta} 
	\\
	& + \frac{1}{2} \left(C_{\alpha \beta} + C_{\alpha \beta} \right).
	\end{split}
	\end{equation}
	where $S_{\alpha \beta} = \frac{1}{2} \left(\partial_\alpha u_\beta + \partial_\beta u_\alpha \right)$, and we have defined:
	\begin{equation}
	\Phi' = \frac{3}{5}(1 - \alpha_R)( \rho_R + \Lambda \rho_B ) - \frac{1}{3} \rho .
	\end{equation}
	To obtain $\Pi_{\alpha \beta}^{(1)}$ in Eq.~(\ref{Pi_1}), multiply Eq.~(\ref{oe}) by $c_{i \alpha} c_{i \beta}$, sum, 
	substitute the definition of $f_i^{(0)}$ (Eq.~(\ref{equ_equ})), use Eqs.~(\ref{Apro1}) to introduce eigenvalue $\lambda_3$ 
	and, crucially, use Eq.~(\ref{equ_advection}) (i.e. the model kinematics) to eliminate terms like $\frac{\partial}{\partial t_0}(2 \phi_1+4\phi_2)$, \cite{Burgin}. 
	That is, the form of $\Pi_{\alpha \beta}^{(1)}$ in Eq.~(\ref{Pi_1}) relies on the fact that $\rho^N$ is a material invariant, on the shortest timescales.
	Use the viscous stress definition, Eqs.~(\ref{equ_viscous_stress}), (\ref{Pi_1}), and simplify:
	\begin{equation}
	\begin{split}
	\label{VS}
	\frac{\sigma_{\alpha \beta}'}{\delta_t} &= -\frac{1}{2\lambda_3} \left(C_{\alpha \beta} + C_{\beta \alpha} \right) + \frac{2}{3} \left(\frac{1}{\lambda_3} - \frac{1}{2}\right) \rho S_{\alpha \beta} 
	\\
	&+ \left(\frac{1}{\lambda_3} - \frac{1}{2}\right) \left[u_\alpha \left(G_\beta - \partial_\beta \Phi' \right) + u_\beta \left(G_\alpha - \partial_\alpha \Phi'\right) \right] 
	\\
	&- \left(\frac{1}{\lambda_3} - \frac{1}{2}\right) \left[u_\gamma \partial_\gamma \Phi' - \frac{1}{3} \rho \partial_\gamma u_\gamma \right] \delta_{\alpha \beta}.
	\end{split}
	\end{equation}
	The discrepancy between the desired result (a term in $\rho S_{\alpha \beta}$ ) and Eq.~(\ref{VS}) defines an error:
	\begin{eqnarray}
	E_{\alpha \beta} & = & -\frac{1}{2\lambda_3} \left(C_{\alpha \beta} + C_{\beta \alpha}\right) \\ \nonumber
				 & + & \left( \frac{1}{\lambda_3} - \frac{1}{2} \right) \left[u_\alpha \left(G_\beta - \partial_\beta  \Phi' \right) + u_\beta \left(G_\alpha - \partial_\alpha \Phi' \right) \right] \\ \nonumber
				 & - & \left( \frac{1}{\lambda_3} - \frac{1}{2} \right) \left[ u_\gamma \partial_\gamma \Phi' - \frac{1}{3} \rho \partial_\gamma u_\gamma\right] \delta_{\alpha \beta}.
	\end{eqnarray}
	Therefore, we make the following choice for $C_{\alpha \beta} $:
	\begin{equation}
	\begin{split}
	\label{Correction}
	C_{\alpha \beta} &= \left(1 -  \frac{\lambda_3}{2} \right) \left[u_\alpha \left(G_\beta - \partial_\beta \Phi' \right) + u_\beta \left(G_\alpha - \partial_\alpha \Phi' \right) \right] 
	\\
	&- \left(1 -  \frac{\lambda_3}{2} \right)  \left[u_\gamma \partial_\gamma \Phi' - \frac{1}{3} \rho \partial_\gamma u_\gamma \right]\delta_{\alpha\beta},
	\end{split}
	\end{equation}
	whence, from Eq.~(\ref{VS}), $\sigma_{\alpha \beta}' = \frac{2}{3} \left(\frac{1}{\lambda_3} - \frac{1}{2}\right) \rho S_{\alpha \beta}\delta_t$, so our model's kinematic viscosity
	is $\nu = \frac{1}{6} \left(\frac{2}{\lambda_3} - 1\right)$. Further, we are also now able to write a local expression for the viscous stress in our large density difference model from 
	Eqs.~(\ref{equ_Pis}) and (\ref{equ_viscous_stress}):
	\begin{equation}
	\sigma_{ \alpha \beta}' = - \left( 1 - \frac{\lambda_3}{2} \right) \sum_i f^{(1)} c_{ i \alpha} c_{i \beta} + \frac{\delta_t}{4}\left(  C_{\alpha \beta } + C_{\beta \alpha }\right), 
	\end{equation}
	with $C_{\alpha \beta}$ defined in Eq.~(\ref{Correction}).
	With this $C_{\alpha \beta}$, source $ F_i $ in Eq.~(\ref{evo}) is partitioned into a term responsible for correcting for density gradients associated with component changes, $F_{1i}$, and one for the interface force $F_{2i}$:
	\begin{equation}
	\label{Source}
	F_i = F_{1i} + F_{2i},
	\end{equation}
	where, conforming to 	Eq.~(\ref{equ_source1}):
	\begin{eqnarray}
	\label{equ_T_identity}
	 T_{\alpha \beta} & = &  \frac{1}{2 c_s^4}\left(1 - \frac{\lambda_3}{2} \right) \bigg\{ \frac{1}{3} \rho \partial_\gamma u_\gamma \delta_{\alpha\beta}  \\ 
	\nonumber
	& & -\left(u_\alpha \partial_\beta \Phi' + u_\beta  \partial_\alpha \Phi' + u_\gamma \partial_\gamma \Phi'\delta_{\alpha\beta} \right) \bigg\}.
	\end{eqnarray}
        and Eq.~(\ref{equ_source2}) gives $F_{2i}$ (which differs dignificantly from that Guo et. al. derive, for a uniform density LBGK).
	
	We now turn to the modal projection.
	We encapsulate the collision source term within the evolution of the modes defined in Table~\ref{tabA1}. 
	In doing so, the advantages of Dellar's MRT scheme are preserved and we shall be able to produce a collision step which is particularly implementable. 
	Define matrix left row eigenvectors:
	\begin{equation}
	\label{eq42}
	\mathbf{M}\equiv \left( \mathbf{h}^{(0)}, \mathbf{h}^{(1)}, \cdots, \mathbf{h}^{(8)} \right)^T,
	\end{equation}
	such that:
	\begin{equation}
	\label{equ_key}
	\mathbf{m}=\mathbf{M}\ \mathbf{f} = \left( \rho, \rho u_x, \rho u_y, \sigma_{xx}, \sigma_{yy}, \sigma_{xy}, N, J_x, J_y \right)^T.
	\end{equation}
	Eq.~(\ref{evo}) is left multiplied by $\mathbf{M}$:
	\begin{equation}
	\label{evo-pro}
	\mathbf{M \ f^+} = \mathbf{M \ f} + \mathbf{M \ A \ M^{-1}} \left( \mathbf{M \ f^{(0)}} -\mathbf{M \ f} \right) +\mathbf{M \ F},
	\end{equation}
	where $\mathbf{F}$ denotes the column vector with elements $F_i$, and $\mathbf{f}$, $\mathbf{f}^+$ and $\mathbf{f}^{(0)}$ are column vectors. 
	 $\mathbf{h^{(n)}}$ are left (row) eigenvectors of $\mathbf{A}$, hence $\mathbf{M\ A}=\mathbf{ \Lambda \ M}$, or:
	\begin{equation}
	\label{eq45}
	\mathbf{ \Lambda} = \mathbf{M\ A \ M^{-1}}, \quad \mathbf{\Lambda}= diag (\lambda_0,\ \lambda_1,\ \cdots \ \lambda_8),
	\end{equation}
	where $\lambda_0 = \lambda_1 = \lambda_2 = 0$, $\lambda_3 = \lambda_4 = \lambda_5$ and $\lambda_7 = \lambda_8$. Therefore Eq.~(\ref{evo-pro}) may be written:
	\begin{equation}
	\label{38}
	m^{(p)+} = m^{(p)} + \lambda_p \left( m^{(0) (p)} - m^{(p)} \right) + S^{(p)}, \quad p = 0,1,..,(Q-1),
	\end{equation}
	the projection of $F_i$ in Eq.~(\ref{evo}) giving modal source:
	\begin{equation}
	S^{(p)} = \sum_{j} M_{pj} F_j.
	\end{equation}
	
	We aim to determine the $S^{(p)}$, then to invert the transformation in Eq.~(\ref{evo-pro}). 
	Bearing in mind the structure of $\mathbf{M}$, it is straightforward to show:
	\begin{eqnarray}
	\nonumber
	S^{(0)}=\mathbf{h}^{(0)} \cdot \mathbf{F}&=& \sum_i h_i^{(0)} F_i = A \equiv 0, \\
	\nonumber
	S^{(1)}= \mathbf{h}^{(1)} \cdot \mathbf{F} &=& \sum_i h_i^{(1)} F_i = \sum_i c_{ix} F_i = n F_x \delta_t, \\
	\nonumber
	S^{(2)}=\mathbf{h}^{(2)} \cdot \mathbf{F}&=& \sum_i h_i^{(2)} F_i = \sum_i c_{iy} F_i = n F_y \delta_t, \\
	\nonumber
	S^{(3)}= \mathbf{h}^{(3)} \cdot \mathbf{F}&=& \sum_i h_i^{(3)} F_i = \sum_i c_{ix}^2 F_i = C_{xx}, \\
	\nonumber
	S^{(4)}=\mathbf{h}^{(4)} \cdot \mathbf{F}&=& \sum_i h_i^{(4)} F_i = \sum_i c_{iy}^2 F_i = C_{yy}, \\
	\nonumber
	S^{(5)}=\mathbf{h}^{(5)} \cdot \mathbf{F}&=& \sum_i h_i^{(5)} F_i = \sum_i c_{ix}c_{iy} F_i \\
	\nonumber
	&=& \frac{1}{2} (C_{xy}+C_{yx}), \\
	\nonumber
	S^{(6)}=\mathbf{h}^{(6)} \cdot \mathbf{F}&=& \sum_i h_i^{(6)} F_i = \sum_i g_i F_i \\
	\nonumber
	&=& - \frac{1}{2}\left(C_{xx} + C_{yy}\right),\\
	\nonumber
	S^{(7)}=\mathbf{h}^{(7)} \cdot \mathbf{F}&=& \sum_i h_i^{(7)} F_i = \sum_i g_i c_{ix} F_i = 0,\\
	\nonumber
	S^{(8)}= \mathbf{h}^{(8)} \cdot \mathbf{F}&=& \sum_i h_i^{(8)} F_i = \sum_i g_i c_{iy} F_i = 0.
	\end{eqnarray}
	Note, source $F_i$ has no projection onto the non-hydrodynamic modes $N$, $J_x$, $J_y$. Projections of $\mathbf{f^{(0)}}$ are also required:
	\begin{eqnarray}
	\nonumber
	\mathbf{h}^{(0)} \cdot \mathbf{f}^{(0)}&=& \sum_i h_i^{(0)} f_i^{(0)} = \sum_i 1_i f_i^{(0)}  =\rho, \\
	\nonumber
	\mathbf{h}^{(1)} \cdot \mathbf{f}^{(0)}&=& \sum_i h_i^{(1)} f_i^{(0)} = \sum_i c_{ix} f_i^{(0)} = \rho u_x, \\
	\nonumber
	\mathbf{h}^{(2)} \cdot \mathbf{f}^{(0)} &=& \sum_i h_i^{(2)} f_i^{(0)} = \sum_i c_{iy} f_i^{(0)} = \rho u_y, \\
	\nonumber
	\mathbf{h}^{(3)} \cdot \mathbf{f}^{(0)}&=& \sum_i h_i^{(3)} f_i^{(0)} = \sum_i c_{ix}^2 f_i^{(0)} = \Pi_{xx}^{(0)} \\
	\nonumber
	\mathbf{h}^{(4)} \cdot \mathbf{f}^{(0)}&=& \sum_i h_i^{(4)} f_i^{(0)} = \sum_i c_{iy}^2 f_i^{(0)} = \Pi_{yy}^{(0)}, \\
	\nonumber
	\mathbf{h}^{(5)} \cdot \mathbf{f}^{(0)}&=& \sum_i h_i^{(5)} f_i^{(0)} = \sum_i c_{ix}c_{iy} f_i^{(0)} = \Pi_{xy}^{(0)}, \\
	\nonumber
	\mathbf{h}^{(6)} \cdot \mathbf{f}^{(0)}&=& \sum_i h_i^{(6)} f_i^{(0)} = \sum_i g_i f_i^{(0)} \\
	\nonumber &=& \frac{9}{5} \alpha_R \rho_R + \frac{9}{5} \alpha_B \rho_B - \frac{4}{5} \rho ,\\
	\nonumber
	\mathbf{h}^{(7)} \cdot \mathbf{f}^{(0)}&=& \sum_i h_i^{(7)} f_i^{(0)} = \sum_i g_i c_{ix} f_i^{(0)} = 0,\\
	\nonumber
	\mathbf{h}^{(8)} \cdot \mathbf{f}^{(0)}&=& \sum_i h_i^{(8)} f_i^{(0)} = \sum_i g_i c_{iy} f_i^{(0)} = 0.
	\end{eqnarray}
	We now find from Eq.~(\ref{38}) and Tab.~\ref{tabA1} the following ``forced'' modal evolution equations:
	\begin{eqnarray}
	\label{moev1} & i= 0:\ \ \ \ \ \ \ &\rho^+=\rho, \\ 
	& i= 1: \ \ \ \ \ \ \  &(\rho u_x)^+=\rho u_x +n F_x \delta_t, \\ 
	& i= 2:  \ \ \ \ \ \ \ &(\rho u_y)^+=\rho u_y +n F_y \delta_t, \\ 
	\nonumber & i= 3 \cdots 5:\   &(\Pi_{\alpha\beta})^+ = \Pi_{\alpha\beta} - \lambda_3 \left( \Pi_{\alpha\beta} -\Pi^{(0)}_{\alpha\beta} \right)\\
	& \ \ \ \ \ \ \ \ \ +& \frac{\delta_t}{2} (C_{\alpha\beta} +C_{\beta\alpha}), \\
	& i= 6: \ \ \ \ \ \ \   & N^+ = N - \lambda_6 N, \\  
	\label{moev7} & i= 7,\ 8:  \ \ \ & J_\alpha^+ = J_\alpha - \lambda_7 J_\alpha, 
	\end{eqnarray}
	where subscripts $\alpha,\ \beta = x,y$. We note the simple form of the relaxation equations for $m^{(6)} \cdots m^{(8)}$, i.e.  $N$, $J_x$, $J_y$, 
	which for $\lambda_6=\lambda_7=1$, reduce to $N^+=J_x^+=J_y^+=0$. 
	
	Having found the forced evolution equations for all the modes $m^{(p)}$, we turn at last to the inversion, from mode space, directly to obtain the distribution function. 
	We define column vectors $\mathbf{k}^{(p)}$:
	\begin{eqnarray}
	k_i^{(0)} &=& 2w_i -\frac{3}{2} w_i \left( c_{ix}^2+c_{iy}^2 \right), \\
	k_i^{(1)} &=& 3w_i c_{ix}, \\
	k_i^{(2)} &=& 3w_i c_{iy}, \\
	k_i^{(3)} &=& \frac{9}{2} w_i c_{ix}^2 -\frac{3}{2} w_i, \\
	k_i^{(4)} &=& \frac{9}{2} w_i c_{iy}^2 -\frac{3}{2} w_i, \\
	k_i^{(5)} &=& 9w_i c_{ix} c_{iy}, \\
	k_i^{(6)} &=& \frac{1}{4} g_i w_i, \\
	k_i^{(7)} &=& \frac{3}{8} g_i w_i c_{ix}, \\
	k_i^{(8)} &=& \frac{3}{8} g_i w_i c_{iy}.
	\end{eqnarray}
	It is straightforward, using the isotropy lattice properties expressed in Eqs.~(\ref{Apro1} to show the $\mathbf{k}^{(p)}$s have the property 
	$\mathbf{h}^{(p)} \cdot \mathbf{k}^{(p')} = \delta_{pp'}$ and hence:
	\begin{equation}
	\label{eq84}
	\mathbf{M^{-1}} = \left( \mathbf{k}^{(0)}, \mathbf{k}^{(1)}, \cdots, \mathbf{k}^{(8)} \right).
	\end{equation}
	Having found $\mathbf{M}^{-1}$, it is now possible to reconstruct a post-collision distribution function vector $\mathbf{f}^+= \mathbf{M}^{-1} \ \mathbf{m}^+$ 
	which, on appeal to Eq.~(\ref{38}) gives:
	\begin{eqnarray}
	\nonumber
	f_i^+&=&(M)_{ij}^{-1}\ m_j^+\\
	\nonumber
	&=& w_i \Bigg\{  \bigg[ 2-\frac{3}{2}\left( c_{ix}^2+c_{iy}^2 \right) \bigg] \rho \\
	\nonumber
	& & \ \ \ \ \ + 3 \left( (\rho u_x)^+ c_{ix} + (\rho u_y)^+ c_{iy}\right)   \\
	\nonumber
	& & \ \ \ \ \ + \frac{9}{2} \left( \Pi_{xx}^+ c_{ix}^2 +2\Pi_{xy}^+ c_{ix}c_{iy} +\Pi_{yy}^+ c_{iy}^2 \right)\\
	\nonumber
	& & \ \ \ \ \  -\frac{3}{2} \left(\Pi_{xx}^+ + \Pi_{yy}^+\right)\\
	\nonumber
	& & \ \ \ \ \ + \frac{1}{4} g_i N^+ + \frac{3}{8} g_i \left( J_x^+ c_{ix} + J_y^+ c_{iy} \right) \Bigg\},
	\end{eqnarray} 
	with the $(\rho u_x)^+$, $(\rho u_y)^+$, $\rho^+$, $\Pi_{xx}^+$, $\Pi_{xy}^+$, $\Pi_{yy}^+$, $N^+$, $J_x^+$ and $J_y^+$ determined in Eqs.~(\ref{moev1}-\ref{moev7}) above. 
	Species or color is finally re-allocated according to Eq.~(\ref{eq_re_color}).
	
	Sources $S^{(p)}$ which rely on kinetic equation source term $F_i$  
	may require spatial numerical derivatives of e.g. density. Computation of such derivatives is important for scheme stability and accuracy.
	The latter is enhanced by use of higher order stencils, as discussed below.
	\section{High order lattice stencils}
	\label{sec_stencils}
	It is possible to exploit lattice tensor isotropy, to develop non-compact stencils of any chosen order of accuracy for first gradient quantities. 
	Thampi et al. have given a similar treatment of this essential approach \cite{SucciStencils} but based around the other gradient quantities (the Laplacian). 
	Consider a scalar function denoted $f$. No confusion with the color-blind distribution function, $f_i$, should arise from use of this notation. 
	A multi-variate Taylor expansion, on the lattice, of function $f(\mathbf{r})$ may be written:
	$f(\mathbf{r} + N \mathbf{c}_i) = f(\mathbf{r} ) + \sum_{n=1}^{\infty} \frac{N^n}{n!}\left( \mathbf{c}\cdot \nabla \right)^n f $.
	Taking moments of this expansion with $w_i c_{i x}$ and appealing to lattice properties (\ref{equ_moments}), we straightforwardly obtain: 
	\begin{equation}
	\label{equ_HOstencil}
	\sum_{i} w_i f(\mathbf{r} + N \mathbf{c}_i) c_{ix}= \frac{N}{3}\frac{\partial f}{\partial x} + \sum_{n=2}^{\infty} \frac{N^{(2n-1)}}{(2n+1)!}E_{(2n-1)},
	\end{equation}
	where $N \in \mathbb Z^+,$ and we define the m$^{th}$ error term:
	\begin{equation}
	E_{(m)} = \left( \sum_{i=1}^Q  w_i c_{ix} c_{i \alpha_1} c_{i \alpha_2}.. c_{i \alpha_m} \right) \left( \frac{ \partial^m f } { \partial x_{\alpha_1} \partial x_{\alpha_2} ...\partial x_{\alpha_m}} \right).
	\end{equation}
	We so not need expressions for the $E_{(m)}$ to eliminate them.
	Let us obtain a non-compact stencil for $\frac{\partial f}{\partial x }$, correct to (say) fifth order, using straightforward linear algebra methods.
	Take $N = 1,2,3$ in Eq.~(\ref{equ_HOstencil}) and truncate each equation at $n>3$, to obtain three equations (one for each choice of $N$).
	These three equations may be written as follows:
	\begin{gather}
	\begin{bmatrix} \sum_{i} w_i f(\mathbf{r} +\mathbf{c}_i) c_{ix}  \\  \sum_{i} w_i f(\mathbf{r} + 2\mathbf{c}_i) c_{ix} \\ \sum_{i} w_i f(\mathbf{r} +3\mathbf{c}_i) c_{ix}
	\end{bmatrix}
	=
	\begin{bmatrix}
	\frac{1^1}{1!} & \frac{1^3}{3!} & \frac{1^5}{5!}  \\ \frac{2^1}{1!} & \frac{2^3}{3!} & \frac{2^3}{5!}   \\\frac{3^1}{1!} & \frac{3^3}{3!} & \frac{3^5}{5!}  \\
	\end{bmatrix}
	\begin{bmatrix}
	\frac{1}{3} \frac{\partial f}{\partial x}    \\ E_{(3)}    \\ E_{(5)}    \\
	\end{bmatrix}.
	\end{gather}
	The inverse matrix of co-efficients, $C_{ij} = \frac{i^{(2j-1)}}{(2j-1)!}$ exists and may be computed. 
	Inverting the above, then, we find an expression for $\frac{\partial f}{\partial x}$ as: 
	\begin{gather}
	\frac{\partial f}{\partial x} =
	\begin{bmatrix}\frac{9}{2} & - \frac{9}{10} & \frac{1}{10} \\
	\end{bmatrix}
	\begin{bmatrix} \sum_{i} w_i f(\mathbf{r} +\mathbf{c}_i) c_{ix}  \\  \sum_{i} w_i f(\mathbf{r} + 2\mathbf{c}_i) c_{ix} \\ \sum_{i} w_i f(\mathbf{r} +3\mathbf{c}_i) c_{ix}.
	\end{bmatrix}
	\end{gather}
	This approach may be adapted to yield expressions for gradients of chosen accuracy. 
	\section{Transient multi-component flows with transverse density stratification}
	\label{sec_test2}
	We consider the semi-analytic, transient flows used in Sec.~(\ref{sec_results}).
	These are, essentially base states of perturbed flows such as those developed by Kao \cite{Kao} and Yih \cite{Yih}, which we obtain, here, by straightforward application of Sturm-Liouville theory.  
	We use similar methodology on two cases of uni-directional, density stratified flow tangent to, first, a flat interface, then, second, a curved interface. 
	We assume the separated fluids have identical shear viscosity, $\eta_1=\eta_2 = \eta$, so the only variation between their 
	kinematic viscosities arises from density.
	
\begin{figure}[ht]
		\begin{center}
			\includegraphics[width=7.0cm]{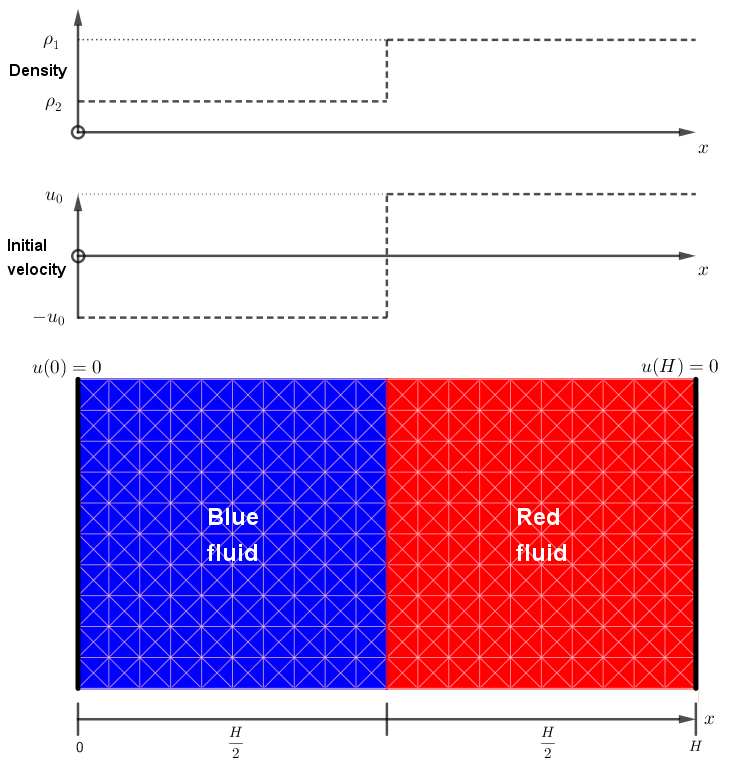}
			\caption{ Schematic. Geometry and initial conditions.
				Fluid is in uni-directional flow, $u(x) \hat{e}_y$. 
				There is translational invariance in the $y$-direction and density stratification,
				with the red (blue) fluid assumed to have density $\rho_1$ ($\rho_2$).
				Initially, the red (blue) fluid moves in the $y$ ($-y$) direction.
			}	
			\label{fig_system}
		\end{center}
	\end{figure}

	Consider flow $u(x,t) \hat{e}_y$, (see Fig.~\ref{fig_system}) with translational invariance in the $y$-direction and no-slip boundaries at $x=0,H$. 
	The flow is modeled as being density stratified with inter-facial boundary conditions introduced as matching conditions on the 
	solution's two pieces. Using the Navier-Stokes equations, the problem is written:
	\begin{eqnarray} 
	\label{eq:Main}
	\nonumber && \rho (x) \frac{\partial  }{\partial t} u(x,t)=  \frac{\partial}{\partial x} \left( \eta  \frac{\partial }{\partial x} u(x,t) \right),\\
	&& u(0)=u(H)=0,
	\end{eqnarray}
	with matching conditions on $u(x,t)$ applied at $x=\frac{H}{2}$ (below).  
	We seek $u(x,t)$, by modal projection on Sturm-Liouville eigenfunctions, $\phi_n$, with eigenvalues, $c_n$ \cite{Arfken}:
	\begin{equation} 
	\label{PDE_Trans}
	u( x, t) = \sum_{n=1}^{\infty} \sigma_n e^{- c_n^2 t}  \phi_n (x),
	\end{equation}
	(where $\sigma_n$ is a constant to be determined), such that:
	\begin{eqnarray} 
	\label{eq:Guess}
	\nonumber
	&& \frac{d}{d x}\left( \eta \frac{d \phi_n(x)}{dx}\right) + c_n^2 \rho(x) \phi_n(x)= 0,\\
	\nonumber
	&& \phi_n(0)=\phi_n(H)=0, \\
	&& \int_0^H \rho(x) \phi_n (x) \phi_m (x) dx = \delta_{nm}.
	\end{eqnarray}

	\( \phi_n \) is obtained piecewise, by solving Eq.~(\ref{eq:Guess}) :
	\begin{align}
	\label{eq: PHI1}
	\phi_n^{(1)} (x) &= A_n^{(1)} \sin \left(\frac{c_n}{\sqrt{\nu_1}} x \right), \ \  x \in \left[ 0, \frac{H}{2} \right), \\
	\label{eq: PHI2}
	\phi_n^{(2)} (x) &= A_n^{(2)} \sin \left(\frac{c_n}{\sqrt{\nu_2}} (x- H) \right), \ \  x \in \left[ \frac{H}{2}, H \right],
	\end{align}
	where $A_n^{(1)}$ and $A_n^{(2)}$ are integration constants and $\nu_i = \frac{\eta}{\rho_i}$, $i =1,2$.
	The kinematic condition ($\phi_n^{(1)} \left(H/2\right) = \phi_n^{(2)} \left(H/2\right)$)
	and the continuous traction condition ( $\left[ \phi_n^{(1)'} \right]_{H/2} = \left[ \phi_n^{(2)'} \right]_{H/2}$ ) 
	provide matching conditions via Eqs.~(\ref{eq: PHI1}, \ref{eq: PHI2})- for non-trivial $A_n^{(1)}$ and $A_n^{(2)}$: 
	\begin{eqnarray}
	\sqrt{\rho_2}  \tan\left(\frac{ c_n H }{2\sqrt{\nu_1}} \right) + \sqrt{ \rho_1}  \tan\left(\frac{c_n H}{2 \sqrt{\nu_2}} \right) = 0.
	 \end{eqnarray}
	By treating $c_n$ as a continuous variable, this equation was solved using the Newton-Raphson method.
	Having thus determined the $c_n$, use the kinematic condition and the ortho-normality property, (Eq.~(\ref{eq:Guess})), to show:
	\begin{eqnarray}  
	\label{eq:FindAn1p}
	\nonumber
	A_n^{(1)}&=& 2 \bigg[ \rho_1  \left( H  - \frac{\sqrt{\nu_1}}{ c_n} \sin \left( \frac{c_n H}{\sqrt{\nu_1}} \right) \right) \\
	\nonumber
	&& + \frac{ \sin^2 \left( \frac{c_n H}{2 \sqrt{\nu_1}}\right)}{\sin^2 \left( \frac{c_n H}{2 \sqrt{\nu_2}}\right)}    
	\rho_2  \left( H  - \frac{\sqrt{\nu_2}}{ c_n} \sin \left( \frac{c_n H}{\sqrt{\nu_2}} \right) \right) \bigg]^{-1/2},\\
	\end{eqnarray}
	\begin{equation}  
	\label{eq:FindAn2}
	A_n^{(2)} = -A_n^{(1)} \frac{ \sin \left( \frac{c_n H}{2 \sqrt{\nu_1}}\right)}{\sin\left( \frac{c_n H}{2 \sqrt{\nu_2}}\right)}.
	\end{equation}
	Finally, we determine the $\sigma_n$, using the initial conditions: 
	\begin{align}
	\nonumber
	&\sigma_n = A_n^{(1)} \frac{u_0}{c_n} \bigg[ \rho_1  \sqrt{\nu_1}  \left( 1 -\cos \left(\frac{c_n H}{2\sqrt{\nu_1} }  \right) \right) \\
	&- \rho_2\sqrt{\nu_2} \frac{ \sin \left( \frac{c_n H}{2 \sqrt{\nu_1}}\right)}{\sin\left( \frac{c_n H}{2 \sqrt{\nu_2}}\right)} \left(1 - \cos\left(\frac{c_n H}{2 \sqrt{\nu_2}}  \right) \right) \bigg].     
	\label{sigma_m_1} 
	\end{align}
	In summary, our transient flow's solution is defined by Eqs.~(\ref{PDE_Trans}), (\ref{eq:FindAn1p}) and ( \ref{sigma_m_1}). 
	
\begin{figure}[ht]
		\begin{center}
			\includegraphics[width=7.0cm]{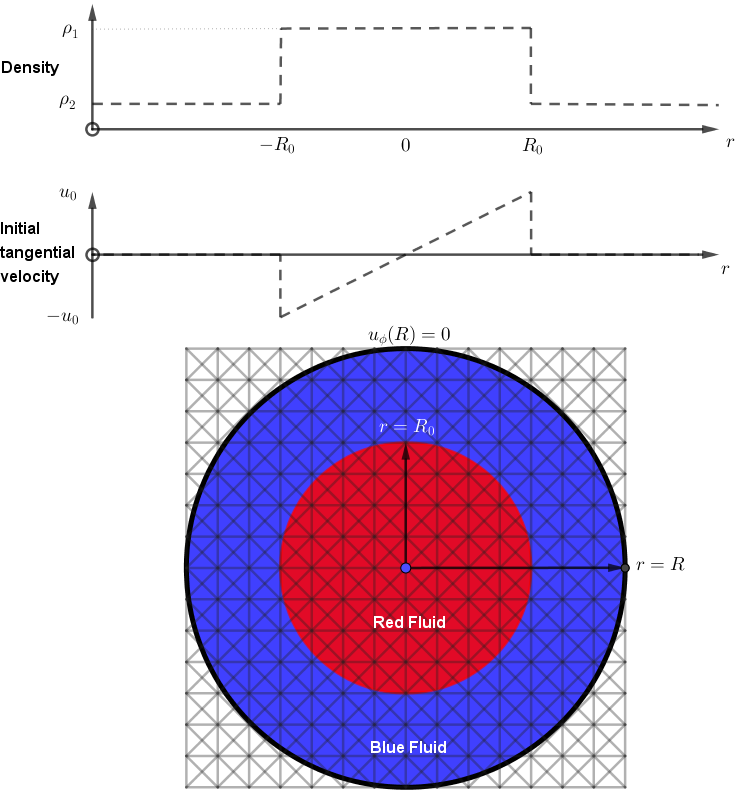}
			\caption{ Schematic. Geometry and initial conditions.
				A weakly compressible fluid is in rotational flow $u_{\phi}(r) \hat{e}_{\phi}$. 
				$r$ represents a ``transverse" co-ordinate.
				A no-slip boundary is located at $r=[R]$. 
				There is transverse density stratification with the red (blue) fluid having density $\rho_1$ ($\rho_2$).
				Initially, the red (blue) fluid moves in the $\hat{e}_{\phi}$ (is at rest).
			}	
			\label{fig_radial_system}
		\end{center}
	\end{figure}

	We consider, now, flow in the axially symmetric geometry of Fig.~(\ref{fig_radial_system}).
	The initial condition is $u_\phi (r, 0)= r (\Theta(r) - \Theta(r-R_0))$, the only non-zero strain rate is $\epsilon_{r \phi} = \frac{1}{2} \left[ r \frac{\partial}{\partial r} \left(  \frac{1}{r}u_\phi(r,t) \right) \right]$
	and the fluid stress divergence is $\nabla \cdot \sigma = \frac{1}{r^2} \frac{\partial}{\partial r} \left( r^2 \sigma_{r \phi} \right)$.
	The fluids are Newtonian, with $ \sigma_{r \phi} = 2 \eta \epsilon_{r \phi} $. 
	From the Navier-Stokes equations therefore:
	\begin{eqnarray}
	\label{eq_N-S-E}
	&& \rho(r) \frac{\partial}{\partial t} u_\phi (r,t) = \frac{1}{r^2} \frac{\partial}{\partial r} \left( r^3 \eta \frac{\partial}{\partial r} \left( \frac{1}{r} u_\phi(r,t)\right) \right), \\ \nonumber
	&& u_{\phi}(0,t) = u_{\phi}(R,t) = 0, \\ \nonumber
	&& p_2 = p_1 + \frac{\sigma}{R_0},
	\end{eqnarray}
	with matching conditions applied at $r=R_0$.    
	Let $ u_\phi (r,t) = \sum_{n=1}^\infty \sigma_n e^{- c_n^2 t} \phi_n (r)$, this time using a Sturm-Liouville eigenspectrum such that:
	\begin{eqnarray}
	\label{equ_transfomed_eigen_eq}
	&& \frac{d}{dr} \left( r \frac{d}{dr} \phi_n(r) \right) - \frac{1}{r} \phi_n(r) = - c_n^2 \left( \frac{r}{\nu }\right) \phi_n(r), \\ \nonumber
	&& \phi_n(0) = \phi_n(R)=0, \\ \nonumber
	&&\int_0^R   r  \left( \nu(r) \right)^{-1} \phi_n (r) \phi_m (r) = \delta_{nm}.
	\end{eqnarray}
	Above, we have used an integrating factor to reach Sturm-Liouville form and the weight function
	is $w(r) = \frac{r}{\nu } = \frac{r \rho(r)}{\eta} = r \nu (r)^{-1}$, \cite{Arfken}. $\phi_n(r)$ is obtained by solving Eq.~(\ref{equ_transfomed_eigen_eq}) (Bessel's equation with $n=1$):
	\begin{align}
	\label{equ_phi_radial}
	\phi_n^{(1)} (r) & = A_n^{(1)} J_1 \left( \frac{c_n r}{\sqrt{\nu_1}}\right), \ \  r \in \left[ 0, R_0 \right), \\
	\phi_n^{(2)} (r) &= A_n^{(2)}  \left( J_1 \left( \frac{c_n r}{\sqrt{\nu_2}} \right) -B_n^{(2)}  Y_1 \left( \frac{c_n r}{\sqrt{\nu_2}} \right)  \right), \ \  r \in \left[R_0, R \right],
	\end{align}
	with $B_n^{(2)} =  \frac{J_1 \left( \frac{c_n R}{\sqrt{\nu_2}} \right)}{Y_1 \left( \frac{c_n R}{\sqrt{\nu_2}} \right)}$.
	We determine eigenvalues, $c_n$, using the kinematic and continuous traction conditions as
	$\phi_n^{(1)}(R_0) = \phi_n^{(2)}(R_0)$ and $ \epsilon_{r \phi}^{(1)}|_{R_0} = \epsilon_{r \phi}^{(2)}|_{R_0}$ respectively.
	These provide matching conditions on $\phi_n(r)$ with a non-trivial solution provided $f \left( \frac{c_n R_0}{\sqrt{\nu_1}} \right)=0$, where:
	\begin{align}
	\label{equ_characteristic_equation}
	\nonumber & &  f(X) = -k_a J_1(X) \Bigg[ J_0 \left( k_a X \right) -J_2 \left( k_a X \right) \ \ \ \ \ \ \ \ \ \ \ \ \ \ \ \
	\\
	\nonumber & & - \frac{J_1 \left(k_a k_b X \right)}{Y_1 \left( k_a k_b X \right)} Y_0 \left(k_a X \right) + \frac{J_1 \left(k_a k_b X \right)}{Y_1 \left( k_a k_b X \right)} Y_2 \left( k_a X \right)\Bigg]  
	\\
	& &+ \left[  J_1 \left(k_a X \right) - \frac{J_1 \left( k_a k_b X \right)}{Y_1 \left(k_a k_b X \right)} Y_1 \left(k_a X \right) \right] \left[  J_0(X) - J_2(X) \right].
	\end{align}
	Above, $k_a = \sqrt{\frac{\nu_1}{\nu_2}}$, $k_b = \frac{R}{R_0}$. 
	Eigenvalues $c_n$ were again obtained using Newton-Raphson iteration. 
	Given a set of $c_n$, we can now write:
	\begin{eqnarray}
	\label{equ_an1}
	\nonumber A_n^{(1)} = && \Bigg[ \frac{1}{\nu_1^2} \int_0^{R_0} r \phi_n^{(1)}\phi_n^{(1)} dr \\
	\nonumber && \ \  + \frac{1}{\nu_2^2} \frac{ J_1 \left( {\frac{c_n R_0}{\sqrt{\nu_1}}}\right)}{ J_1 \left( {\frac{c_n R_0}{\sqrt{\nu_2}}}\right)  - B_n^{(2)}(R_0) J_1 \left({\frac{c_n R_0}{\sqrt{\nu_2}}}\right) } \\
	&& \ \ \ \ \ \ \ \ \times \int_{R_0}^R r \phi_n^{(2)} \phi_n^{(2)} dr\Bigg] ^{- \frac{1}{2}}
	\end{eqnarray}
	which was obtained using Simpson's rule. Also:
	\begin{equation}
	\label{equ_an2}
	A_n^{(2)} = A_n^{(1)} \frac{ J_1 \left( {\frac{c_n R_0}{\sqrt{\nu_1}}}\right)}{ J_1 \left( {\frac{c_n R_0}{\sqrt{\nu_2}}}\right)  - B_n^{(2)}(R_0)  J_1 \left( {\frac{c_n R_0}{\sqrt{\nu_2}}}\right) }. 
	\end{equation}
	Integration constants, $\sigma_n$, are determined using initial and ortho-normality conditions on $\phi_n$, as:
	\begin{equation}
	\label{equ_sigma_n_ortho_2}
	\sigma_n = A_n^{(1)} \frac{\rho_1} {\eta}\int_0^{R_0} r^2 J_1 \left( \frac{c_n r}{\sqrt{\nu_1}} \right) dr
	\end{equation}
	which was again evaluated using Simpson's rule. 
	The full transient flow was computed using Eqs.~(D3), (D4), (D5), (\ref{equ_sigma_n_ortho_2}).
	\section{Numerical solution of steady, pressure-driven flow with density stratification }
	\label{sec_sol_Ba_test}
\begin{figure}[ht]
		\begin{center}
			\includegraphics[width=8.8cm]{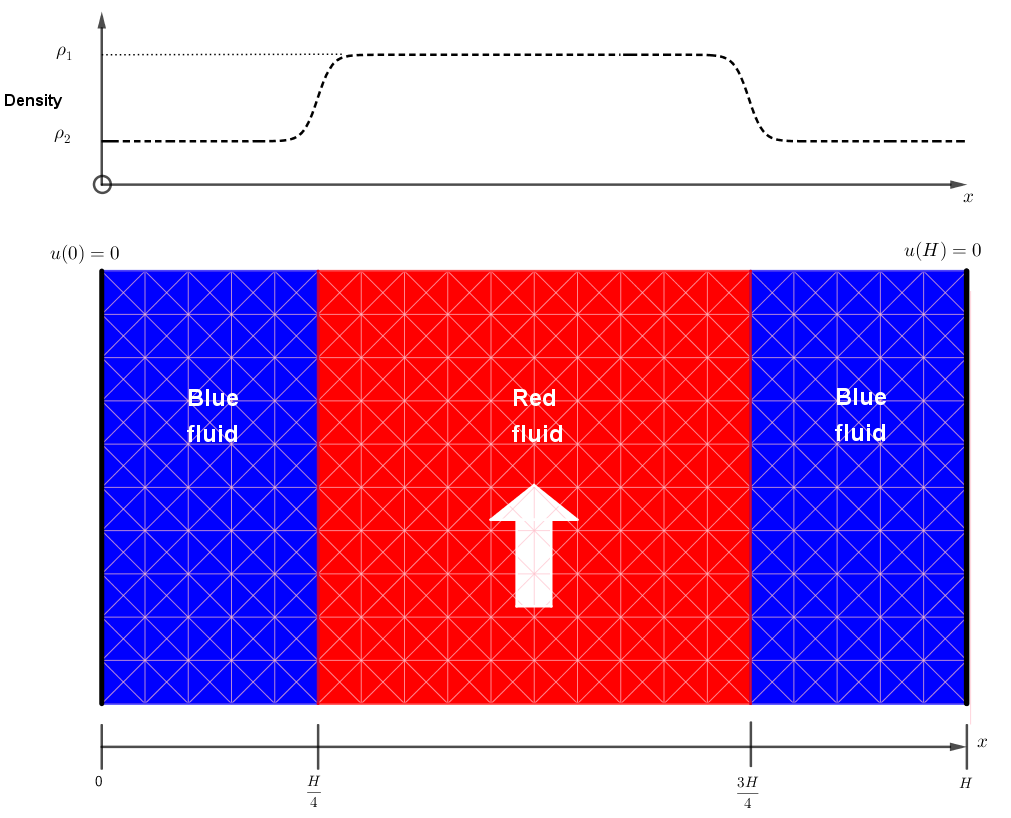}
			\caption{ Schematic representation of the geometry and transverse density stratification used in the pressure gradient (white arrow) driven flow tests.}	
			\label{fig_BaTest}
		\end{center}
	\end{figure}

	Consider the steady-state of the density stratified, uni-directional flow $u(x) \hat{e}_y$, shown in Fig.~\ref{fig_BaTest}, now with a steady pressure gradient $\left( -G \hat{e}_y \right)$,
	a continuous transverse variation of density
	\begin{eqnarray}
	\nonumber
	\rho(x) & = & \frac{1}{2} \left( \rho_{0R} + \rho_{0B}\right) \\ \nonumber
	&  +& \frac{1}{2} \left( \rho_{0R} - \rho_{0B}\right) \tanh \left( \beta \left( x - \frac{H}{4} \right) \right) \Theta \left( \frac{H}{2} - x \right) \\ \nonumber
	&  +& \frac{1}{2} \left( \rho_{0R} - \rho_{0B}\right) \tanh \left( \beta \left( \frac{3H}{4} - x \right) \right) \Theta \left(  x - \frac{H}{2} \right), \\ 
	\label{equ_density}
	\end{eqnarray}
	where $\Theta(x)$ is the Heaviside function.
	For this flow, the Navier-Stokes equation for a weakly compressible lattice fluid and the associated boundary and symmetry conditions are respectively:
	\begin{eqnarray} 
	\label{equ_ODE}
	\nonumber
	&& \frac{d}{d x} \left( \eta(x) \frac{d }{d x} u(x) \right) = G, \quad u(0)=u(H)=0,\\
	&& \left[ \frac{du}{dx} \right]_{H/2}= 0,
	\end{eqnarray}
	where $\rho$ , $u$, and $\eta$ again denote the density, velocity and shear viscosity of the fluid respectively. 
	Shear viscosity, $\eta$, varies continuously with $x$ when the kinematic viscosity, $\nu(\lambda_3)$ = constant, due to the variation in $\rho(x)$ identified in Eq.~(\ref{equ_density}), note. 
	Let $x\leq \frac{H}{2}$. Integrating ordinary differential equation Eq.~(\ref{equ_ODE}) and eliminating the integration constant using the symmetry condition, we have
	\begin{equation}
	\frac{du}{dx} = \frac{G(2x - H)}{2\eta(x)}.
	\end{equation}
	Substituting $\eta(x)=\nu(\lambda_3) \rho(x)$, integrating over range $[0, x]$ with $x < \frac{H}{2}$, using the boundary condition $u(0)=0$ and using a dummy variable, we obtain
	\begin{equation}
	\label{equ_integral}
	u(x) = \frac{G}{2 \nu(\lambda_3)} \int_0^{x} \frac{(2 \alpha -H)}{\rho(\alpha)} d \alpha.
	\end{equation}
	The integral in Eq.~(\ref{equ_integral}) was evaluated numerically, using Simpson's rule, using the expression for density given in Eq.~(\ref{equ_density}). 
	%
	%
	%


	\pagebreak

\end{document}